\documentclass[aps,amsfonts,pre,twocolumn,superscriptaddress,showpacs,eqsecnum]{revtex4-1}
\usepackage{epsfig,amsmath,amssymb,bm,epsf,graphics,psfrag,verbatim,subfigure,color}

\def\lw{w} 
\def\ll{l} 
\def\lh{h} 
\def\numbrace{N_b} 
\def\pbrace{p} 
\def\side{L} 
\def\probm{\mathcal{P}_m} 
\def\numem{N_{em}} 
\def\numcol{N_\textrm{col}} 
\def\nummax{N_M} 
\def\mth{m^\textrm{th}} 
\def\bpc{c} 
\def\probrigid{\mathcal{P}_{\textrm{rigid}}} 
\def\probrigidedge{R} 
\def\numem{N_{em}} 
\def\probbulk{\mathcal{P}_\textrm{bulk}} 
\def\avgfloppy{\left\langle F\right\rangle} 
\def\numfloppy{F} 
\def\numsites{N} 
\def\dim{d} 
\def\compmat{\mathbf{R}} 
\def\nc{N_c} 
\def\rigidthreshold{p_r}

\def\numbraceR{N_{b,r}}
\def\numbraceG{N_{b,g}}

\begin{document}


\title{Rigidity percolation by next-nearest-neighbor braces on 
generic and regular isostatic lattices}

\author{Leyou Zhang}
\affiliation{Department of Physics, University of Michigan, Ann
Arbor, MI 48109, USA}

\author{D. Zeb Rocklin}
\affiliation{Department of Physics, University of Michigan, Ann
Arbor, MI 48109, USA}

\author{Bryan Gin-ge Chen}
\affiliation{Instituut-Lorentz for Theoretical Physics, Leiden University, NL 2333 CA Leiden, The Netherlands}

\author{Xiaoming Mao}
\affiliation{Department of Physics, University of Michigan, Ann
Arbor, MI 48109, USA}

\date{\today}

\begin{abstract}
We study rigidity percolation transitions in two-dimensional central-force 
isostatic lattices, including the square and the kagome lattices, as next-nearest-neighbor bonds (``braces'') are randomly added to the system.  In
particular, we focus on the differences between regular lattices, which are perfectly periodic,
and generic lattices with the same topology of bonds but whose sites
are at random positions in space.  We find that the regular square and kagome lattices exhibit a rigidity percolation transition when the number of braces is $\sim L\ln L$, where $L$ is the linear size of the lattice.  This transition exhibits features of both first order and second order transitions: the whole lattice becomes rigid at the transition, whereas there exists a diverging length scale.  In contrast, we find that the rigidity percolation transition in the generic lattices occur when the number of braces is very close to the number obtained from the Maxwell's law for floppy modes, which is $\sim L$.  The transition in generic lattices is a very sharp first-order-like transition, at which the addition of one brace connects all small rigid regions in the bulk of the lattice, leaving only floppy modes on the edge.  We characterize these transitions using numerical simulations and develop analytic theories capturing each transition.  Our results relate to other interesting problems including jamming and bootstrap percolation.

\end{abstract}

\maketitle

\section{\label{sec:level1}Introduction}

Suppose we build a house or some other mechanical structure
based on a square grid.  Such a structure would be
``shaky'' in the following sense---each pair of adjacent rows or columns of
walls can be sheared by only bending the material at the crossing
points, and this has a much lower energy cost than compression/extension.  If we insert a diagonal brace across a grid
square, we will stabilize the structure by removing a shear mode from
the system, \emph{but how many braces do we need to stabilize the whole structure}?

This type of question belongs to a class of problems known as
``rigidity
percolation''~\cite{Phillips1979,Thorpe1983,Feng1984,Phillips1985,Jacobs1995}.
In a typical rigidity percolation problem, one starts from a stable
lattice, removes bonds randomly so that each bond is present with
probability $p$, and then examines the threshold probability
$\rigidthreshold$ where the structure loses mechanical
stability and finally identifies corresponding scaling laws near this point.  For
example, rigidity percolation in diluted generic triangular lattices
occurs near
$p=\rigidthreshold ^{\textrm{generic triangular}}\simeq 0.6602$ and
exhibits a diverging length scale
\begin{align}\label{EQ:xitrig}
	\xi^{\textrm{generic triangular}} \sim \vert p-\rigidthreshold ^{\textrm{generic triangular}}\vert^{-1.21} ,
\end{align}
that characterizes the size of rigid clusters.  Here ``generic'' means
the sites are not on a perfect periodic lattice, so that rigidity only
depends on the connectivity, as we discuss in detail below.
Interestingly, instead of comprising \emph{a single universality class}, the
nature of the rigidity percolation transition is strongly affected by
the lattice architecture and a rich spectrum of phenomena has been
observed.  Besides the two-dimensional (2D) triangular lattice example
discussed above, in three dimensions (3D)~\cite{chubynsky} and on
complete graphs~\cite{moukarzelfield,Kasiviswanathan,rivoire,Barre2014} 
the rigidity percolation transition is first order.
Moreover, the jamming of frictionless spheres, which can also be viewed
as a version of self-organized rigidity percolation, exhibits
mean-field scaling laws and a jump in coordination number,
and has been characterized as a ``mixed first-and-second-order transition''~\cite{Liu1998,Liu2010,Schwarz2006}. 

A new category of rigidity percolation has been studied on periodic
lattices that are at the verge of mechanical instability (called ``isostatic
lattices'', as discussed below), such as the square lattice discussed
in the first paragraph.  Random addition of next-nearest-neighbor
(NNN) bonds (from now on known as {\em braces}) can remove ``floppy
modes'' (i.e., deformations that do not change the length of any bond)
in these
lattices~\cite{Maxwell1864,Mao2010,Ellenbroek2011,Mao2011a,Lubensky2014,Moukarzel1999}
and thus lead to a rigidity percolation transition.  We will call this
category of rigidity percolation problems ``bracing percolation''.
In particular, the rigidity percolation in a braced generic square lattice has been found to have first-order nature in
Ref.~\cite{Moukarzel1999}.  
In the treatment of that paper, the system was pinned along two \emph{diagonal} edges and free along the other two diagonals. As we will see, the boundary plays a crucial role in the rigidity transition, and our system's open boundaries lead to qualitatively new behavior while confirming the essential order of the transition.

To understand the unique features of bracing percolation, it is useful to review how one determines whether a structure has mechanical stability.  Consider normal modes of a $d$-dimensional system.  
The zero-energy modes of this $\dim$-dimensional system can be divided into $\dim(\dim+1)/2$ rigid-body translations and rotations of the whole system, and $\numfloppy$ floppy modes which involve relative displacements between different parts of the system. J.~C.~Maxwell noted in 1864~\cite{Maxwell1864} that for a system of $\numsites$ particles and $\nc^{(i)}$ \emph{independent} constraints,
\begin{align}\label{EQ:Maxwell}
	F = \dim \numsites - \nc^{(i)} - \frac{\dim(\dim+1)}{2}.
\end{align}
The system then becomes rigid (mechanically stable) when $\numfloppy = 0$, at the \emph{isostatic point}. Under this \emph{Maxwell behavior}, each bond placed in the system eliminates a single floppy mode until the system becomes rigid when the number of constraints is
\begin{eqnarray}\label{eq:maxwellnumber}
\dim \numsites - \frac{\dim(\dim+1)}{2}.
\end{eqnarray}
However, in general and as discussed below, some bonds may be redundant and generate self stresses rather than eliminating floppy modes. This leads to the modified Maxwell relationship~\cite{Calladine1978},
\begin{align}\label{eq:calladine}
	\numfloppy = \dim \numsites - \nc - \frac{\dim(\dim+1)}{2} + S,
\end{align}
where the number of floppy modes depends not only on the number of
constraints $\nc$ but also on $S$, the number of self stresses in the system. These self stresses not only determine rigidity but are determined by it---a self stress occurs when a bond connects two sites in a rigid region of the structure.
This modified Maxwell's rule has been shown to be an index theorem for topological surface modes in systems near isostaticity~\cite{Kane2014}.

Large central-force lattices with coordination number $z=2d$ are
called ``isostatic lattices'' because in the bulk each site has equal
numbers of degrees of freedom and constraints (assuming central-force
nearest bonds only)~\cite{Lubensky2014}~\footnote{In a more rigorous
classification, lattices satisfying $z=2d$ are called ``Maxwell
lattices''  and only those ones with no self stress are called
``isostatic lattices''}.  A finite piece of an isostatic lattice has
$\numfloppy\propto \side^{d-1}$ where $\side$ is
the linear size of the lattice and $\numsites\sim \side^\dim$, because sites on the
boundary have fewer than $2\dim$ bonds.  To make such a finite lattice
stable, one could add exactly $\numfloppy$ braces by making sure that they are
all independent, and thus all floppy modes are eliminated.  We can thus define the ``Maxwell number''  $\nummax$ of a finite isostatic lattice,  the minimum number of braces needed to rigidify the lattice if all braces were added independently.

If the braces are instead added randomly, how many does one typically need for
rigidity?  Studies of
bracing percolation on isostatic lattices address this question.  It
is worth pointing out that unlike other rigidity percolation problems,
the rigidity of isostatic lattices is strongly affected by boundary
conditions because isostatic lattices 
have a sub-extensive number of floppy modes owing only to their boundary.  The above discussion
refers to lattices with open boundary conditions.  Changing to periodic
boundary condition may or may not lift the floppy modes, depending on
the architecture of the lattice~\cite{Sun2012}.

In particular, in Ref.~\cite{Ellenbroek2011} it was shown that the
bracing percolation problem on the regular 2D square lattice, which is an
example of an isostatic lattice, can be mapped into a random-graph
problem (see also \cite{bolkercrapo}).  
Thus exact solutions are possible; it was found that if each brace
is present with a uniform probability $\pbrace$, rigidity percolation
in a (regular) periodic square lattice of size $L\times L$ occurs at 
\begin{align}\label{EQ:prSquare}
	\rigidthreshold^{\textrm{regular square}} = \frac{\ln L}{L}+\mathcal{O}(1/L),
\end{align}
where ``regular'' refers to perfectly periodic lattices (see discussions below).  
At this transition, the probability of a site to be part of the infinite rigid cluster jumps from 0 to 1. On the other hand, the form of $\pbrace_r$ suggests a length scale 
\begin{align}
	\xi^{\textrm{regular square}} \sim \pbrace^{-1} ,
\end{align}
corresponding to a characteristic system size that exhibits with high probability mechanical
stability at a given $\pbrace$.  

Interestingly, this scaling relation for the length scale, together with a characteristic frequency $\omega^*\sim \pbrace$~\cite{Mao2010} agree with corresponding scaling relations observed near jamming~\cite{Silbert2005,Wyart2005}, namely
\begin{align}
\omega^*&\sim\Delta z,\nonumber\\
l^*&\sim \Delta z^{-1},
\end{align}
where $\Delta z = \langle z\rangle-2d$ is the coordination above isostaticity and thus the same as $\pbrace$.  
These scaling relations differ from those observed in randomly diluted triangular lattices [Eq.~\eqref{EQ:xitrig}] but agree with those of randomly braced isostatic lattices.

The above results on square lattices are derived for perfectly periodic square
lattices, in which lattice sites sit on a periodic square grid in
space and bonds in each row or column are
collinear~\cite{Guest2003,Connelly2009,Sun2012}.  However, real
physical lattices invariably have sites displaced slightly from
regular lattice positions, and these displacements profoundly alter
the rigidity of the system.  As pointed out in Ref.~\cite{Jacobs1995},
perfect periodic lattices may exhibit self stress because some bonds
may be redundant because they are parallel to each other, and thus to
study the fundamental physics of rigidity percolation one should
eliminate such redundancy coming from the symmetry of the lattices by
randomizing the positions of the lattice sites.  Floppy modes in these
randomized lattices depend only on the lattice's topology rather than
the positions of the sites \cite{GraverServatiusServatius} and these lattices lack the straight lines which allow stress to be transmitted over long distances without decaying
 and thus may exhibit generic properties of rigidity transitions that depend only on the network's connectivity.
Here, we follow the notion of Ref.~\cite{Jacobs1995} to call the
perfect periodic lattices ``regular'' and the randomized version
``generic''.  The rigidity of the generic lattices can be determined
by a fast algorithm called the ``pebble
game''~\cite{Jacobs1995,Jacobs1996,JacobsHendrickson} which is based
on Laman's theorem for rigidity of graphs~\cite{Laman1970}.  The
result we cite in Eq.~\eqref{EQ:xitrig} was obtained by applying the
pebble game algorithm to generic diluted triangular lattices.  Our
results also use this algorithm to determine the rigidity of generic isostatic lattices.

\begin{figure}[h]
\subfigure[]{
\includegraphics[width=0.35\linewidth]{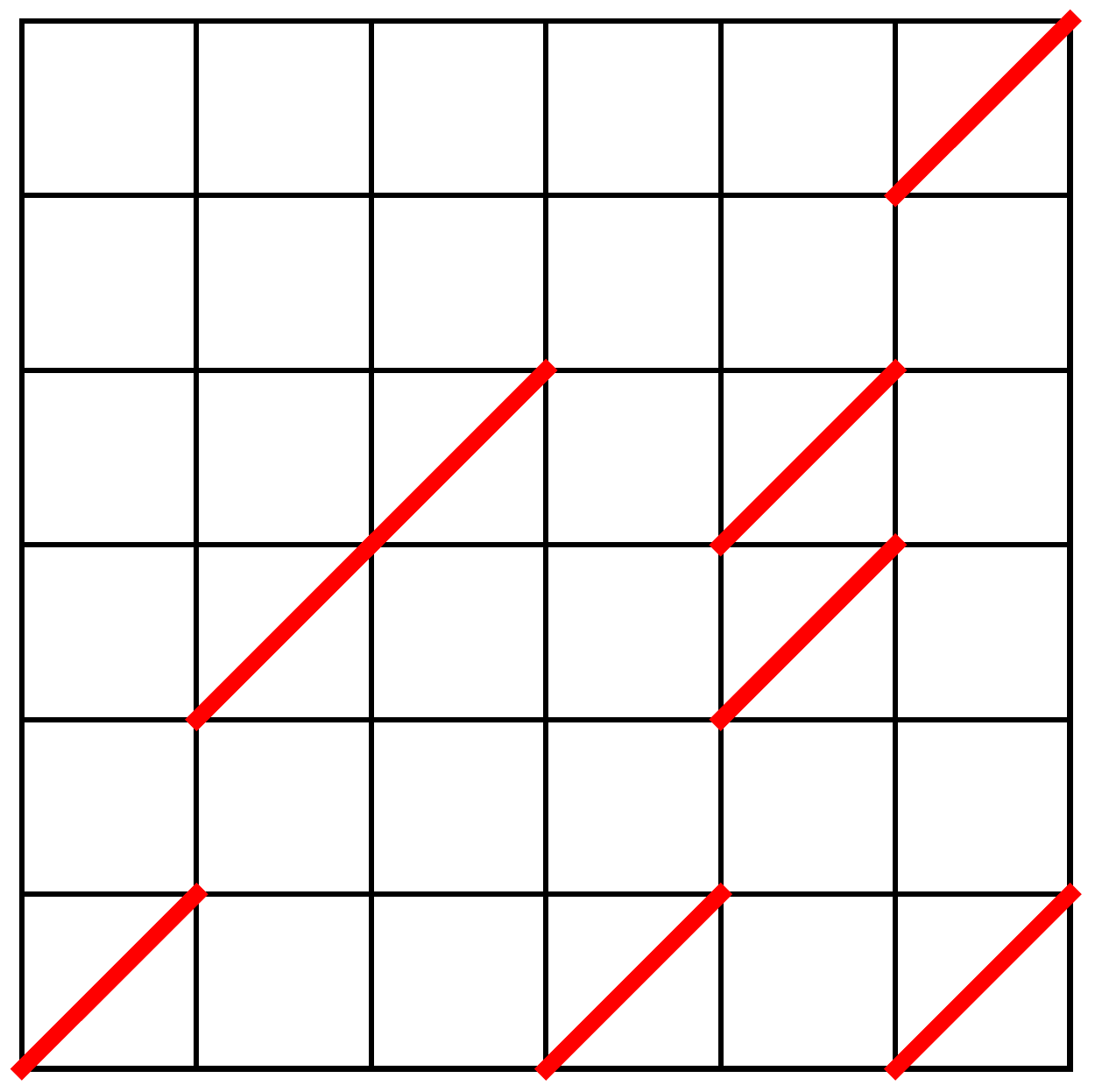}}
\subfigure[]{
\includegraphics[width=0.45\linewidth]{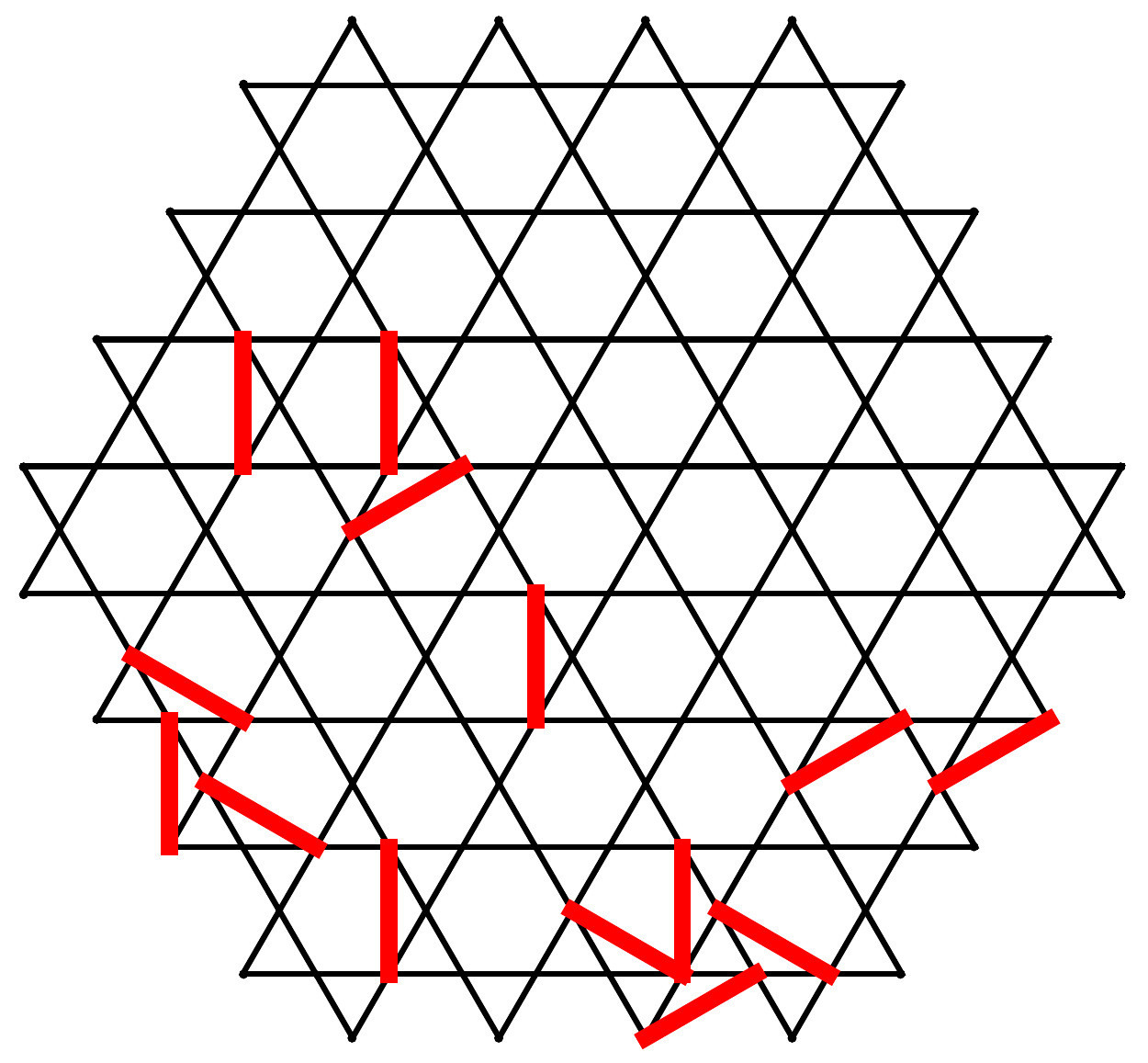}}
\subfigure[]{
\includegraphics[width=0.35\linewidth]{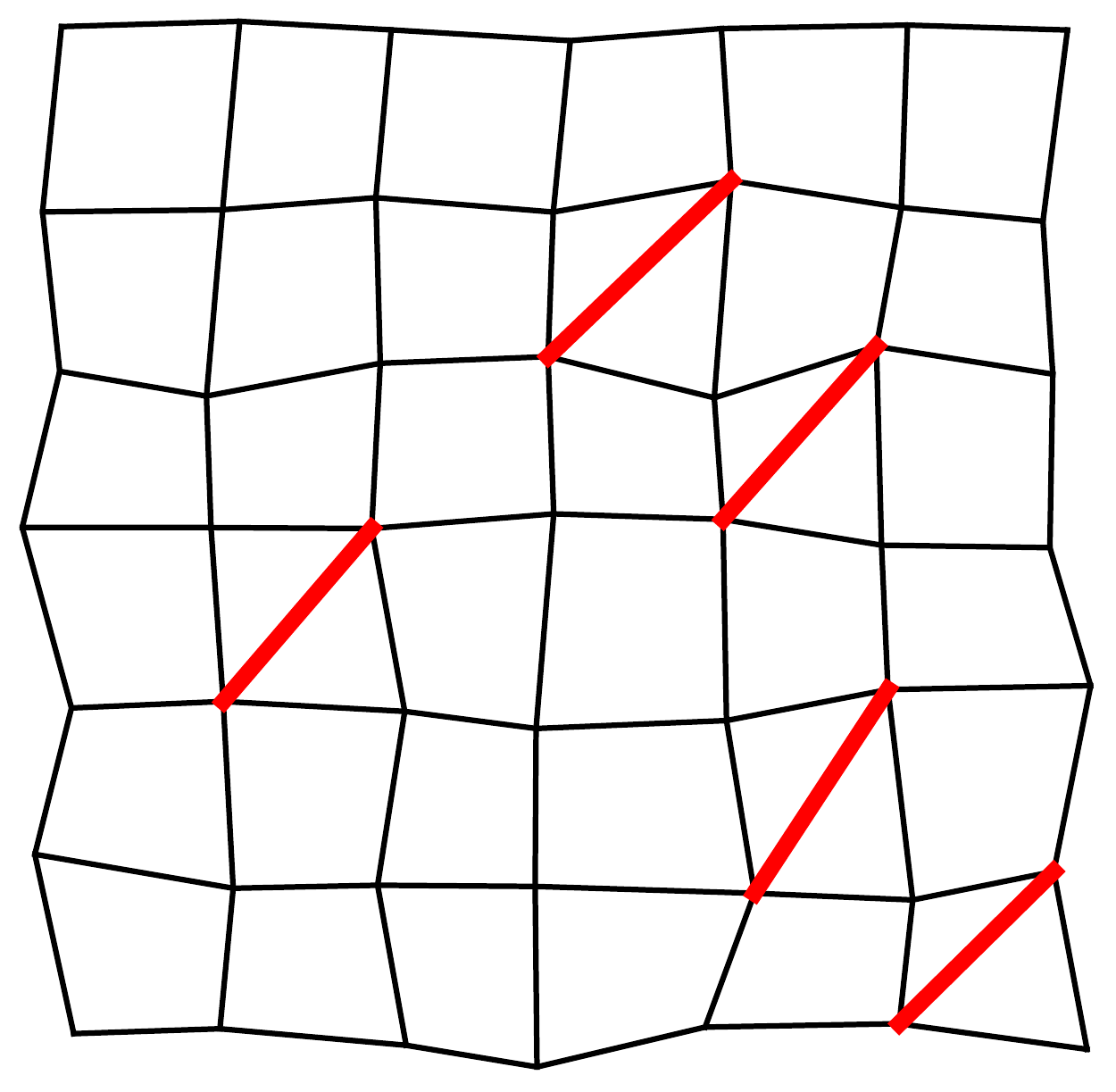}}
\subfigure[]{
\includegraphics[width=0.45\linewidth]{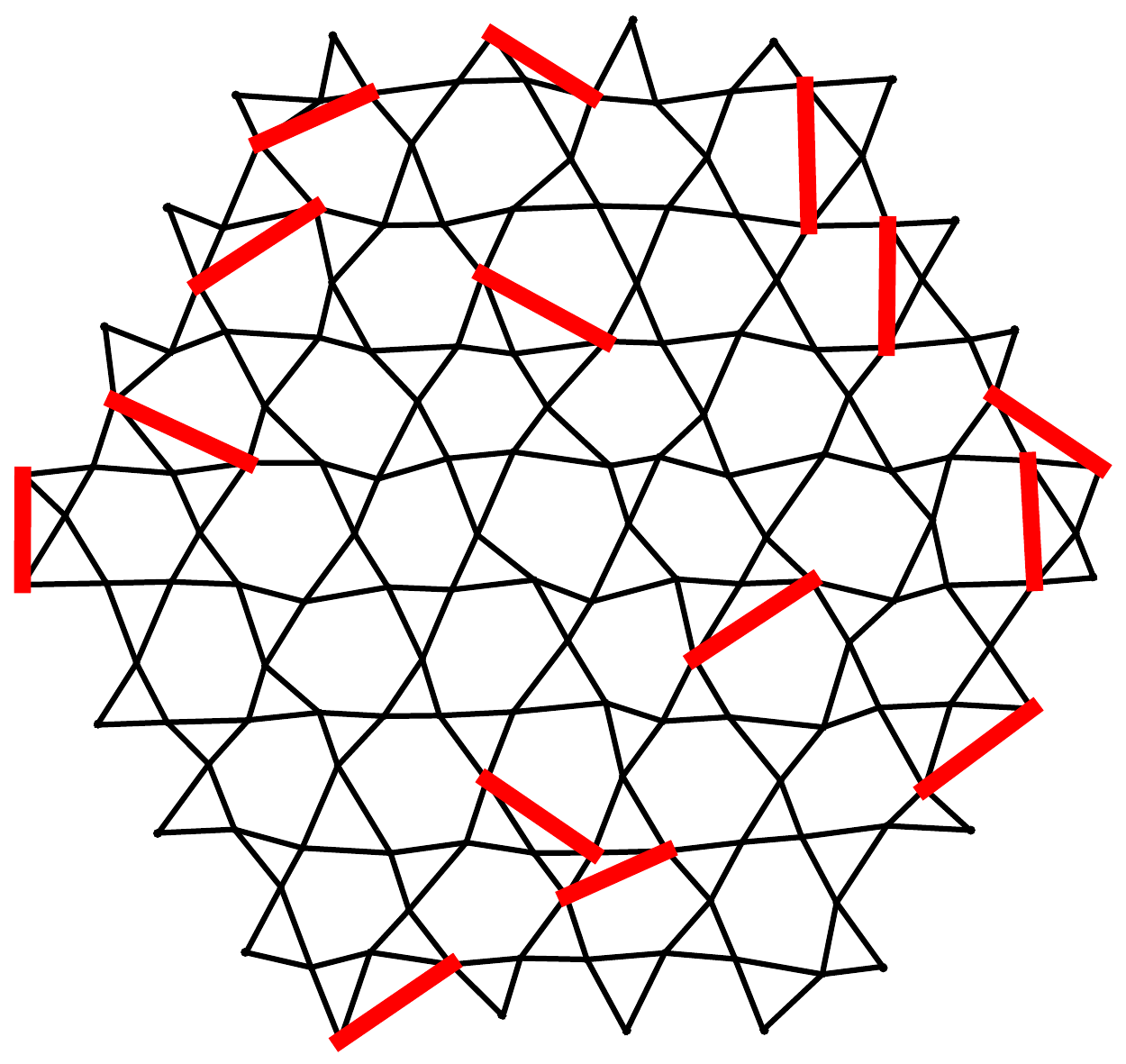}}
\caption{\label{fig:lattice} Illustration of regular square (a),
regular kagome (b), generic square (c) and generic kagome lattices (d)
with nearest-neighbor (NN) bonds (black, thin) and random NNN braces
(red, thick). The square lattices depicted have $\side=7$ and the kagome lattices
have $\side=4$.}
\label{FIG:lattice}
\end{figure}

Studies on the regular square lattice reveal interesting physics that
is intimately related to jamming, but jamming involves random packings
not living on lattices.  It is thus of interest to examine what
changes if one considers generic rather than regular square lattices.

In this Paper, we compare rigidity percolation transitions in braced regular [Fig.~\ref{FIG:lattice}(a,b)] and generic [Fig.~\ref{FIG:lattice}(c,d)] isostatic lattices.   
We investigate how generic and regular square and kagome lattices,
which have $\nummax\sim \mathcal{O}(L)$ floppy modes when no braces
are present, gain rigidity as braces are randomly added to these lattices.  

In particular, instead of having each brace present with a
probability $\pbrace$, we use the total \emph{number} of braces $\numbrace$ as our control parameter.  In other words, we consider the process of randomly adding braces into the lattice one by one.  This is because we find that the rigidity percolation transition in the generic isostatic lattices is an extremely sharp
first order transition, and using fixed $\pbrace$ broadens the
transition window and obscures the sharpness of the transition, as we discuss below.

Our main results are:
\begin{itemize}
\item Both regular square and kagome lattices show a rigidity percolation transition occurring when the number of randomly added braces $\numbrace=\numbraceR\propto \side\, \ln\side$, where $\side$ is the linear size.  This transition shows features of both first order and second order transitions, sharing similarities to many other interesting systems~\cite{Berthier2011,Mezard1999,Liu2010,Schwarz2006,aizenmanlebowitz,holroyd,kineticallyconstrainedreview}.  
In addition, both of these two lattices exhibit another transition at
a lower density of braces, at $\numbraceG\propto \side$, at which the
number of floppy modes shows certain singularities.  Our numerical
results and analytical theory for these phenomena show good agreement.
\item Both the generic square and kagome lattices 
show a very sharp, first-order-like rigidity transition, in contrast
to that of the regular lattices.  In particular, as braces are
randomly added to the generic lattice, floppy modes are eliminated,
following Maxwell's rule [Eq.~\eqref{EQ:Maxwell}] perfectly without
any states of self stress ($S=0$), until the number of braces,
$\numbrace$, becomes close to the total number of floppy modes,
$\nummax-\numbrace \lesssim \mathcal{O}(1)$, when the bulk of the
lattice suddenly rigidifies (which we name \lq\lq bulk rigidity\rq\rq
) as a single brace is added, leaving only $\lesssim
\mathcal{O}(1)$ floppy modes on the \emph{boundary of the lattice}.
After this point, states of self stress start to develop, and within
$\mathcal{O}(1)$ more braces, the whole lattice becomes rigid with
finite probability.  This greatly differs from regular lattices in which
states of self stress develop before an infinite rigid cluster
appears, and it requires $\mathcal{O}(\side\ln\side)$ bonds to
rigidify the system.
\end{itemize}

In Sec.~\ref{SEC:Simulation} we define the models we study and present
our results from numerical simulations, using the \lq\lq pebble
game\rq\rq\ algorithm and direct calculations by evaluating ranks of rigidity matrices.  In Sec.~\ref{SEC:TheoryRegular} we present theoretical results on
the number of floppy modes and the probability of rigidity as
functions of the number of added braces in regular isostatic
lattices.  In Sec.~\ref{SEC:TheoryGeneric} we present theoretical results for generic isostatic lattices including edge modes and statistics of rigidity.  In Sec.~\ref{SEC:CD} we summarize our conclusions and discuss the relation of our work to other studies.

\section{Simulation results}\label{SEC:Simulation}
We begin by defining the family of random spring networks that we
study.

For a square lattice with $\side$ sites per side, initially there are a total of $N=\side^2$ particles,
$2\side^2-2\side$ bonds, and $2\side-3$ floppy modes, so
$\nummax=2\side-3$ (see Fig.~\ref{FIG:lattice}a).    
We add random braces, with only one choice of brace permitted in each
plaquette (bottom-left to upper-right in the figures we show), as a
second such brace (upper-left to bottom-right) is always redundant if the first one is present. There are then $(\side-1)^2$ places braces may be placed.

For the kagome lattice, we study
systems shaped like large hexagons with $\side$ hexagons per side (see Fig.~\ref{FIG:lattice}b). There are $N=9\side^2+3\side$ particles and
initially $18\side^2$ bonds and $\numfloppy = 6 \side -3$ floppy modes
on the kagome lattices, and thus $\nummax=6\side-3$ as well.
We then add random braces.  For kagome lattices, again we allow only
half of the brace positions as the other half are redundant.  Thus for each hexagon, 3 independent bracing positions are allowed. There are then
$9\side^2-3\side$ places braces may be placed.

For these lattices, we generate realizations of disorder by randomly adding $\numbrace$ braces, and mainly evaluate two important quantities: $\probrigid$, denoting the probability the lattice is rigid (i.e., has no floppy modes),
and $\avgfloppy$, denoting the average number of floppy modes, to characterize the rigidity of the lattice.
As we explain below for the two cases of regular and generic lattices, having no floppy modes is equivalent to having a rigid cluster that percolate through the whole system (rigidity percolation).  
We use different computational methods to determine the rigidity of the regular and the generic lattices, as we discuss in detail below.

\subsection{Determining rigidity of regular lattices}\label{Sec:SimuRegu}
The infinitesimal rigidity properties of a spring network can be determined from
the ``rigidity matrix'' (or compatibility
matrix) \cite{Calladine1978}.  This is an $\nc \times \dim \numsites$ matrix $\compmat$ which computes the
vector of bond extensions $e$ from the vector of particle displacements
$u$, i.e. $e=\compmat\cdot u$.  The rank of $\compmat$ gives the number of
independent constraints on the $d \numsites$ degrees of freedom, and so
the dimension of the space of infinitesimal displacements of the particles which do
not stretch any bonds to first order is the dimension of the null
space of $\compmat$,
so that in two dimensions
\begin{eqnarray}
\numfloppy = 2 \numsites - 3 - \text{rank}(\compmat).
\end{eqnarray}

\begin{figure}[h]
\centering
\includegraphics[width=.4\textwidth]{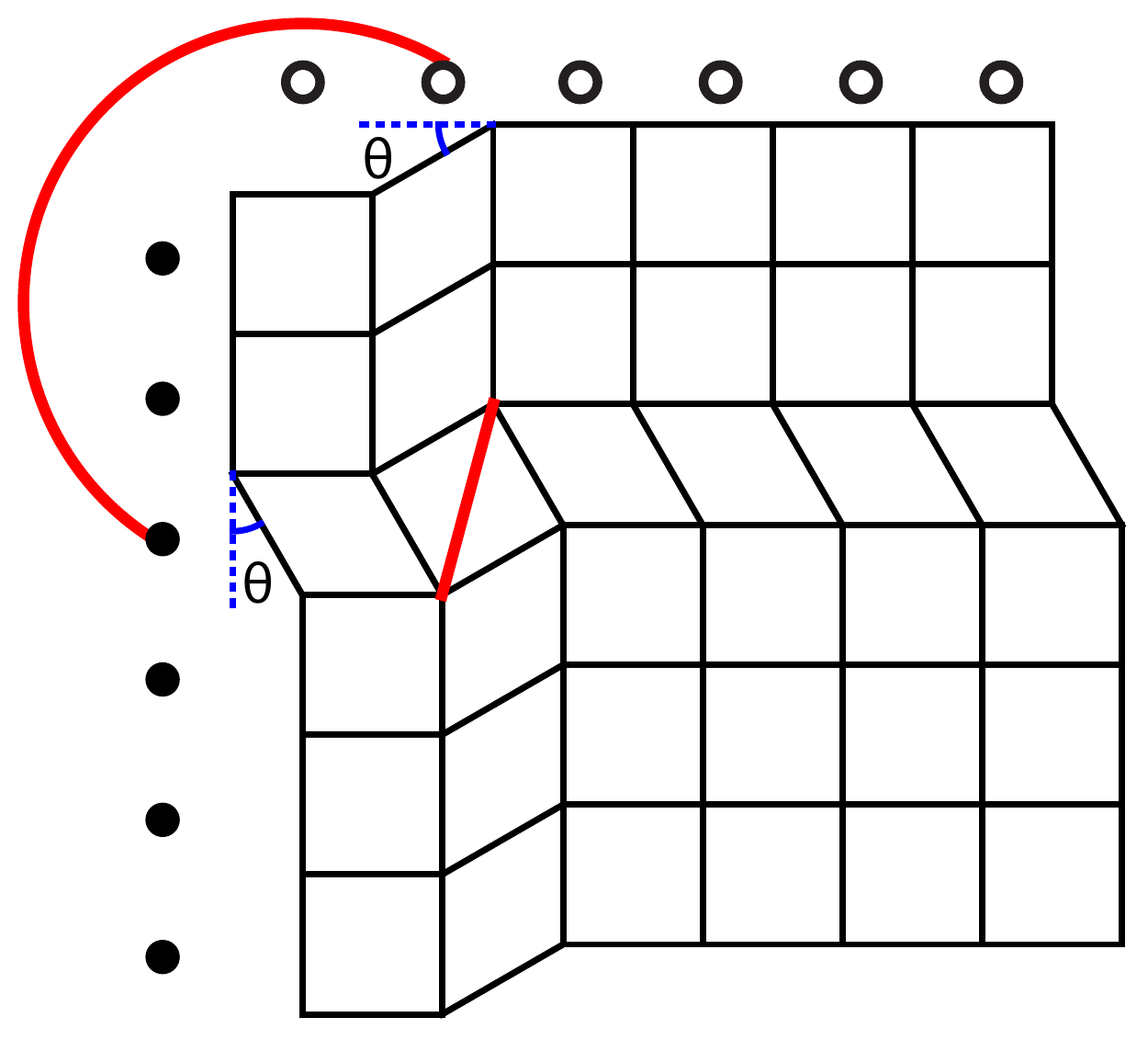}
\caption{
One example of the mapping of braced regular square lattices to bipartite graphs.  In this mapping floppy modes in the ``line-localized'' basis, taking the form of shearing rows (columns), are mapped to black (white) nodes, and braces are mapped to edges (solid red curve) connecting the two types of nodes.  
The deformation of this regular square lattice illustrates how the brace locks the two floppy modes together.}
\label{fig:fmc}
\end{figure}

For the regular lattices, there exist simplified
rigidity matrices, which we call ``braced rigidity matrices''. They
contain $\numbrace$ rows and approximately $\nummax$ columns. 
This simplification from $\mathcal{O}(\side^2)$ columns to $\mathcal{O}(\side)$ columns is
because the floppy modes for the unbraced regular square
lattice and the unbraced regular kagome lattice can
be written in a convenient ``line-localized'' basis. 
The prototype is the regular square lattice with braces, whose
braced rigidity matrix arises from the mapping of its rigidity
properties to bipartite graphs, as shown in Fig.~\ref{fig:fmc} (described in Refs.\
\cite{bolkercrapo,Ellenbroek2011}, see also Sec.\ \ref{SEC:TheoryRegular}).  
Via this mapping, the rigidity of a
set of braces can be determined from the incidence matrix of an
associated bipartite graph \cite{bolkercrapo}, which is the braced
rigidity matrix in this case. This bipartite graph has
$2\side -2 =\nummax+1$ vertices, one for every adjacent pair of rows or columns,
and one edge for every brace. 
For the regular kagome lattice
the braced rigidity matrices are $\numbrace\times(6\side)$
matrices whose construction is outlined in Appendix B. 

Thus, for regular lattices,  
we first calculate $\avgfloppy$ (averaged over random configurations
of 
braces) by calculating the ranks
of the braced rigidity matrices after adding $\numbrace$ independent
and randomly distributed braces. 
We then find $\probrigid$ by
calculating the probability of having $\numfloppy=0$ among all the realizations.  
Because all floppy modes are extended row/column modes, rigidity percolation coincides with $\numfloppy=0$.
For the regular square lattices, we studied $\avgfloppy$ in systems
with linear system sizes $\side=100,200,300$ and averaged over $10^3$
configurations.  To study $\probrigid$ we looked at regular square lattices with $\numfloppy=0$ 
$\side=320,640,1280$.
For the regular kagome lattices, we studied systems with sizes ranging
from $\side=100$ to $\side=800$, averaging over $10^4$ configurations.

\begin{figure}[h]
\subfigure[]{
\includegraphics[width=0.8\linewidth]{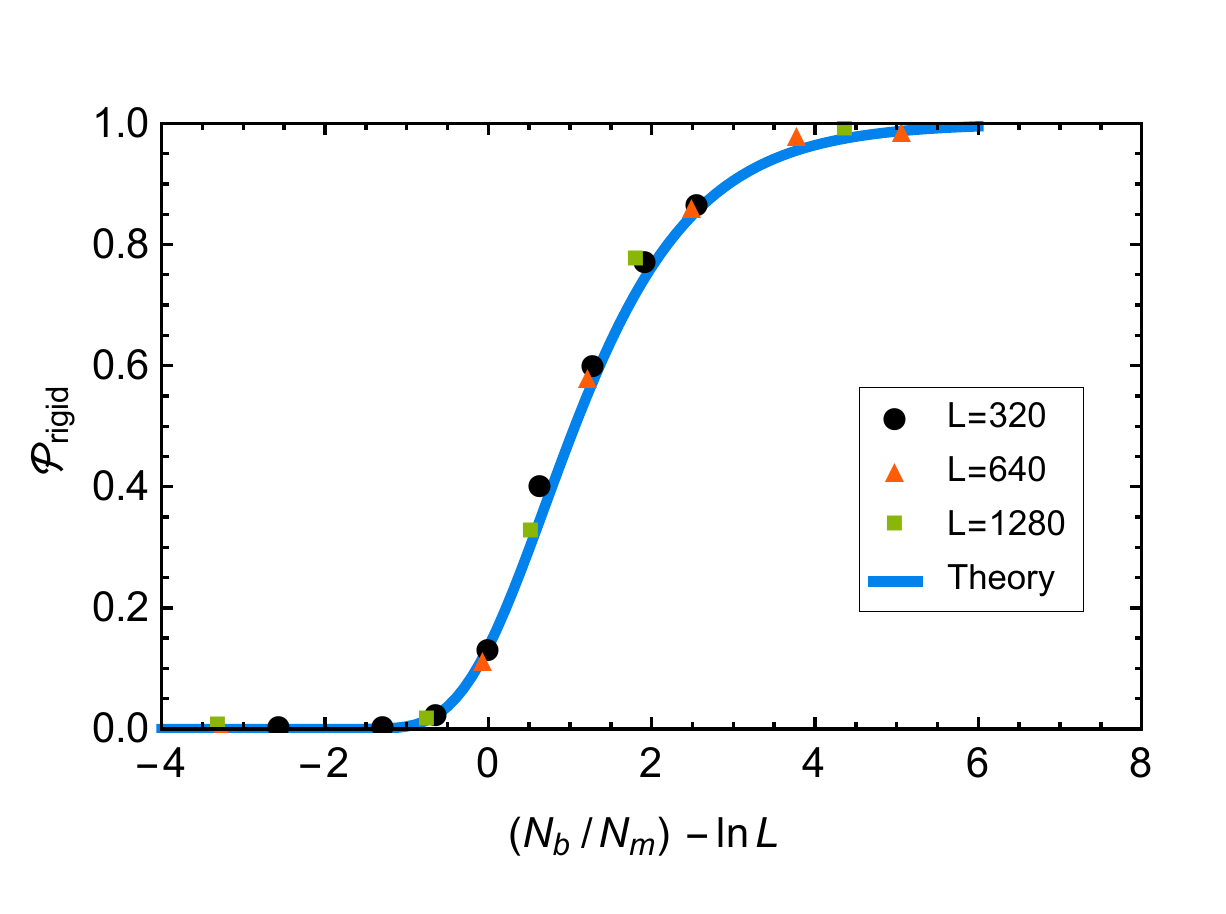}}
\subfigure[]{
\includegraphics[width=0.8\linewidth]{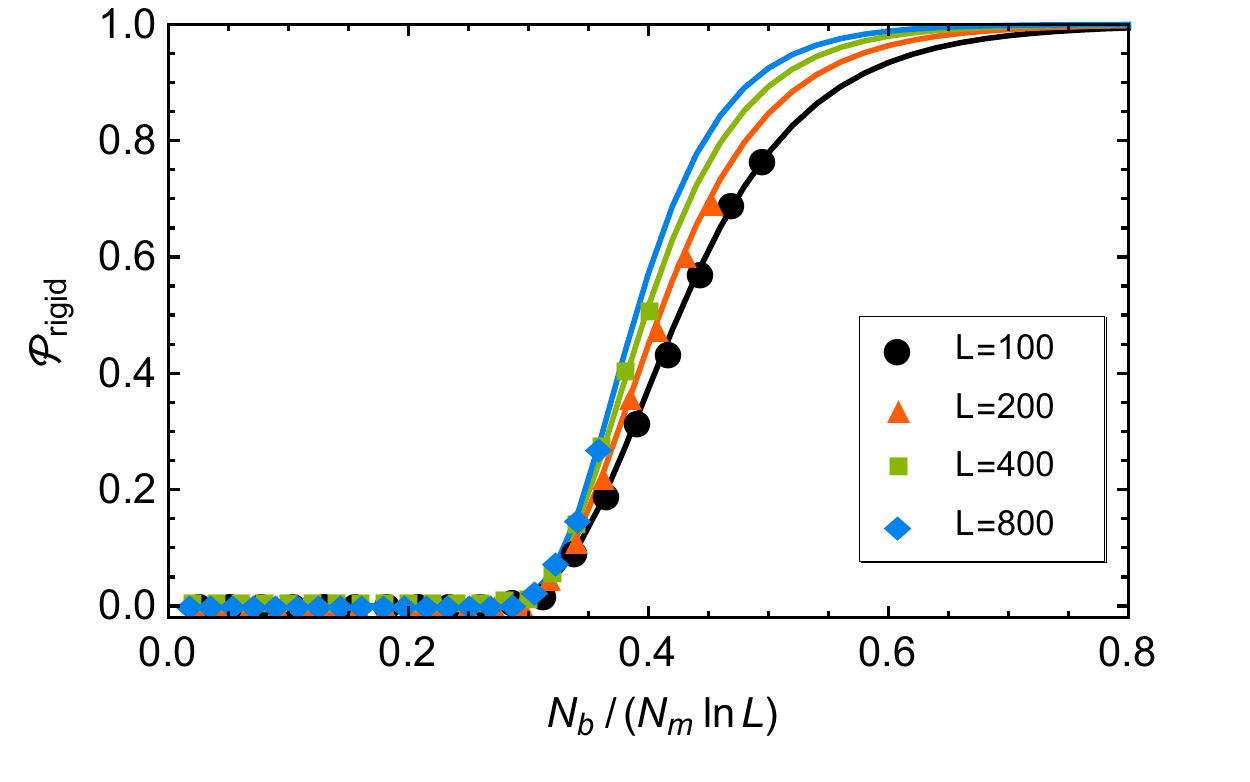}}
\caption{Probability $\probrigid$ for rigidity percolation in braced
 regular square (a) and kagome (b) lattices as a function of scaled number of braces. The 
curve in (a) shows the analytical
asymptotic result in Eq.\ (\ref{eq:regsqrigprob}). 
The curves in (b) show the analytical result of Eq.\ (\ref{eq:regkagrigprob}). }
\label{fig:regular_prob_plots}
\end{figure}

\begin{figure}[h]
\subfigure[]{
\includegraphics[width=0.8\linewidth]{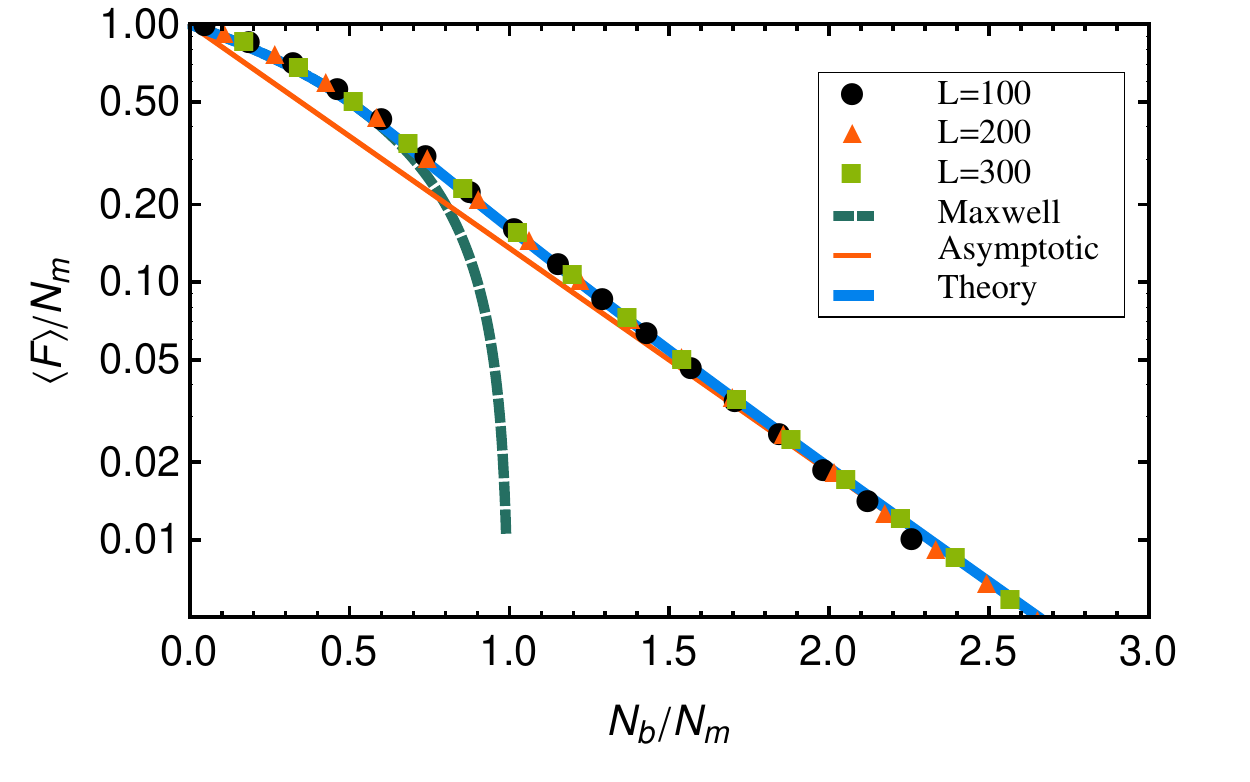}}
\subfigure[]{
\includegraphics[width=0.8\linewidth]{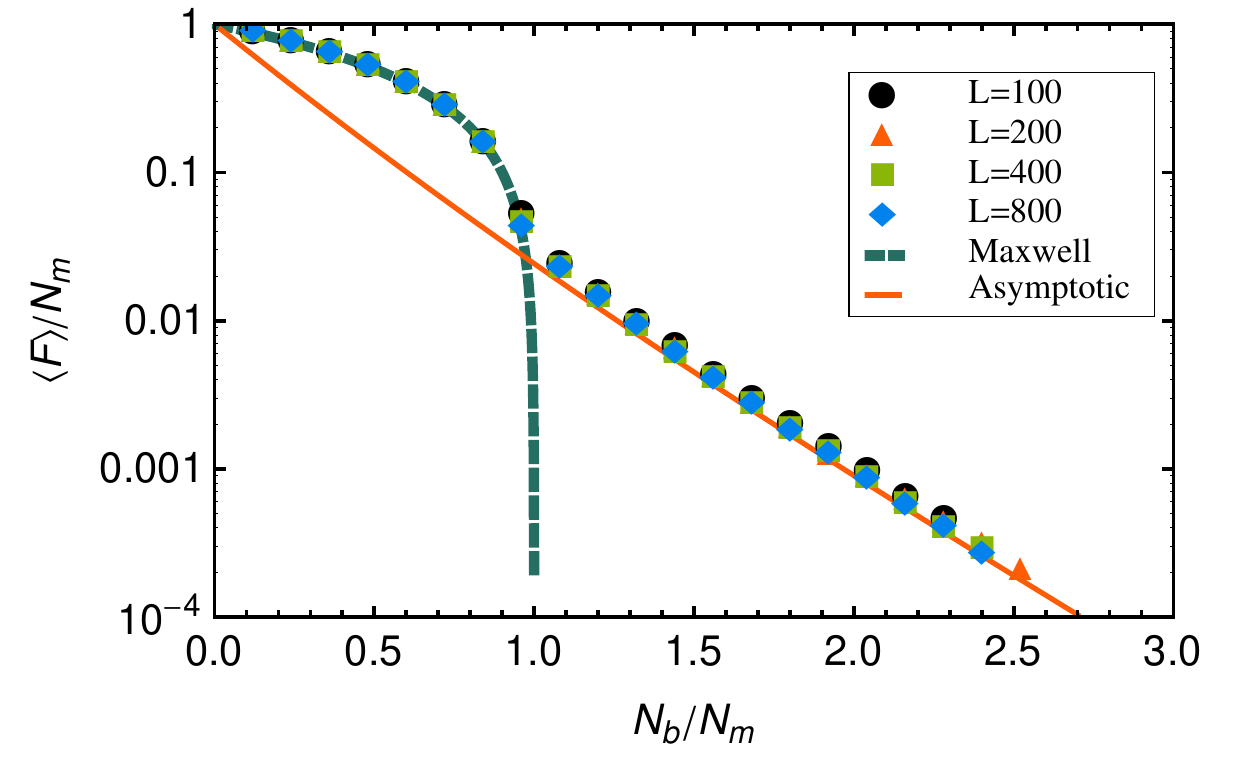}}
\caption{Average number of floppy modes normalized by the Maxwell
number $\avgfloppy/\nummax$ of braced regular square (a)
and kagome (b) lattices as a function of
$\numbrace/\nummax$.  
The asymptotic line in (a) is from Eq.\
(\ref{eq:asymptoticregularsquarefloppy}) and the theory line in (a) is from
Eq.\ (\ref{eq:regularsquarefloppy}).  The asymptotic line in (b) is from Eq.\
(\ref{eq:asymptoticregularkagomefloppy}).
 }
\label{fig:regular_floppy_plots}
\end{figure}

In order to study the spatial pattern of rigidity in regular lattices, we calculate the dynamical matrix~\cite{Lubensky2000} (the null space of which correspond to the floppy modes)
of the regular square and kagome lattices to find the rigid
plaquettes in the regular square lattices and the rigid hexagons in
the regular kagome lattices. 
Self-stressed bonds in the regular square
and kagome lattices are identified by checking whether removing such a bond creates a floppy mode.  

We plot our data for $\probrigid$ for the regular square
and kagome lattices in Fig.~\ref{fig:regular_prob_plots}, data for
$\avgfloppy$ in Fig.\ \ref{fig:regular_floppy_plots}.  Snapshots of
regular square and kagome lattices with various numbers of added
braces are shown in Fig.\ \ref{fig:rigid_regular} with our discussion
on the rigidity transition in generic lattices.

The collapse of the plots of $\probrigid$ at different lattice sizes onto a single line in
Fig.~\ref{fig:regular_prob_plots} shows that the rigidity transitions in regular square and kagome lattices occur at
\begin{align}\label{EQ:NRRegular}
\numbraceR ^{\textrm{regular}}\propto \side \, \ln \side,
\end{align}
in agreement with result of $\pbrace_r^{\textrm{regular square}}$ in Eq.~\eqref{EQ:prSquare} from Ref.~\cite{Ellenbroek2011}, because the probability of having each brace and the (average) total number of braces are related by $\langle \numbrace \rangle = \pbrace (\side-1)^2$.

In addition, our data for $\avgfloppy$ (Fig.~\ref{fig:regular_floppy_plots}) indicate that, before the whole system becomes rigid, there is another transition at 
\begin{align}
\numbraceG^{\textrm{regular}}
	 \propto \nummax \propto \side.
\end{align} 
This is identified from a singularity in $\avgfloppy$, and is
associated to the emergence of a giant cluster of locked floppy modes
(which is not sufficient to rigidify the whole lattice), as detailed in Sec.~\ref{SEC:TheoryRegular}.

\subsection{Determining rigidity of generic lattices}\label{Sec:SimuGene}
For generic lattices, instead of generating rigidity matrices, we
use the ``pebble game'' algorithm, developed
in Refs.~\cite{Jacobs1995,Jacobs1996,JacobsHendrickson} and based on Laman's
theorem~\cite{Laman1970}, to count the number of floppy modes and
identify rigid regions and regions with states of self stress
(overconstrained regions).  We study generic square and kagome
lattices with $\numsites$ ranging from $10^4$ to $1.6\times10^5$.
 For each $\numsites$, we generate $10^4$ realizations of random
distribution of braces. 

For each $\numbrace$, we calculate the average number of floppy modes
$\avgfloppy$ and the probability for the whole
lattice to be rigid $\probrigid$, i.e., to have $\numfloppy=0$.
As with the regular lattices, the last floppy modes of the generic lattices spread across the system, and so rigidity percolation is not achieved until  $\numfloppy=0$.

\begin{figure}[h]
\label{generic_pr}
\centering
\subfigure[]{
\includegraphics[width=0.8\linewidth]{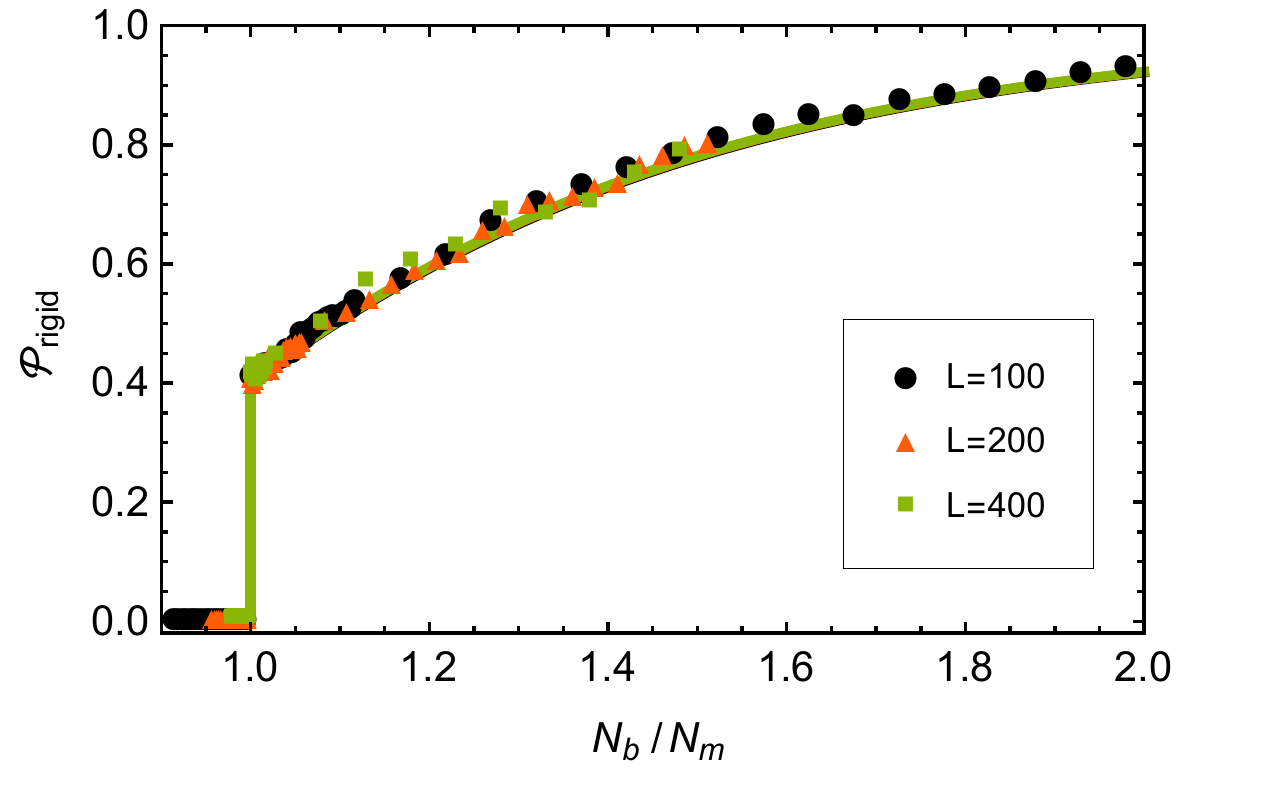}}
\subfigure[]{
\includegraphics[width=0.8\linewidth]{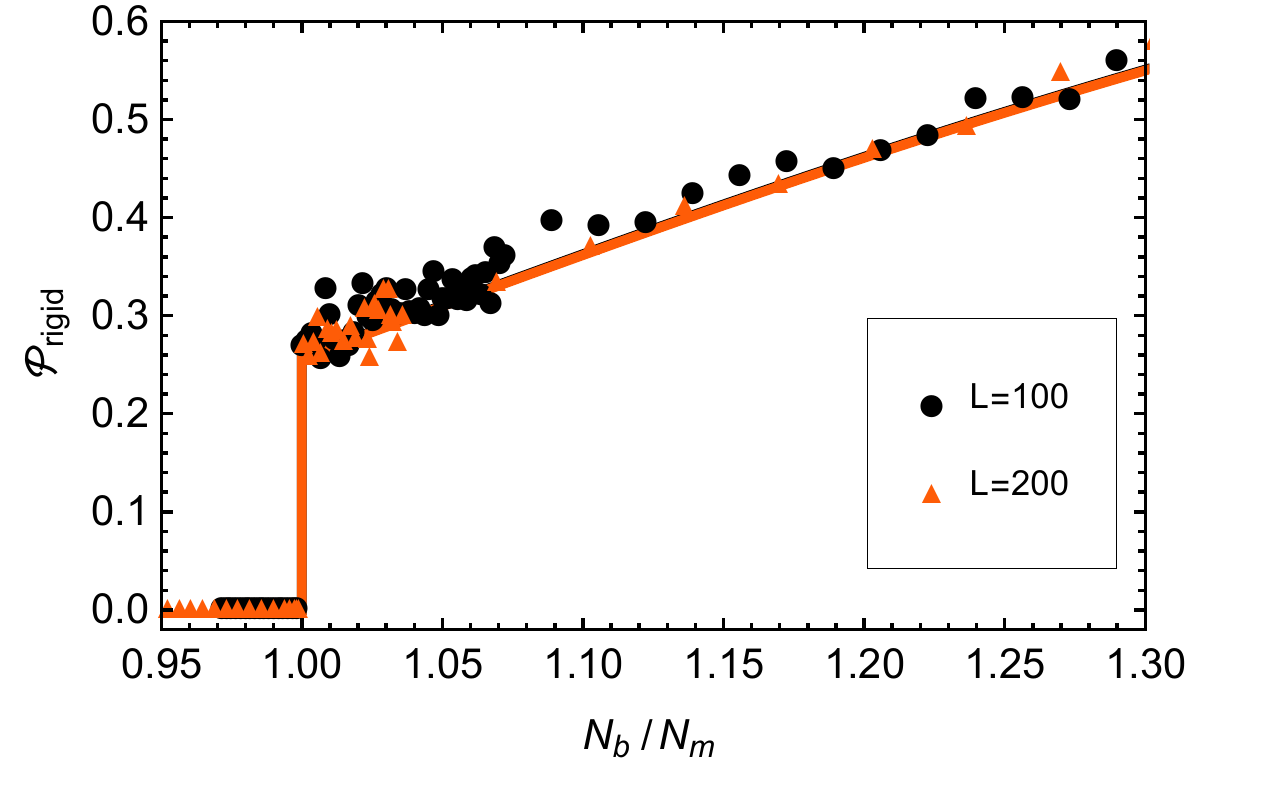}}
\caption{Probability of rigidity percolation $\probrigid$ on (a)
generic square lattices and (b) generic kagome lattices as a function
of  $\numbrace/\nummax$. $\probrigid$ remains zero until
jumping to a finite value at the Maxwell point, indicating a first-order
like transition. The solid lines show the theoretical result of Eq.~(\ref{eq:probrigid}).
\label{fig:pr}}
\end{figure}

\begin{figure}[h]
\label{generic_f}
\subfigure[]{
\includegraphics[width=0.8\linewidth]{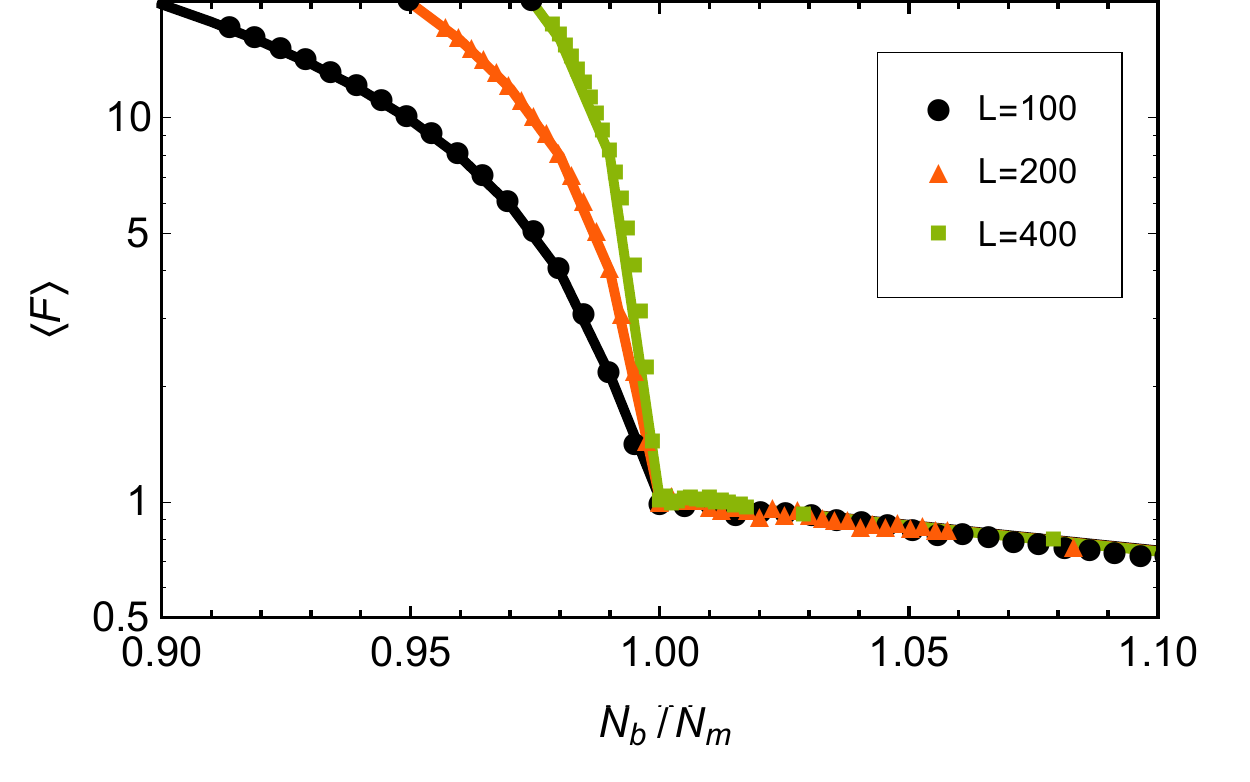}}
\subfigure[]{
\includegraphics[width=0.8\linewidth]{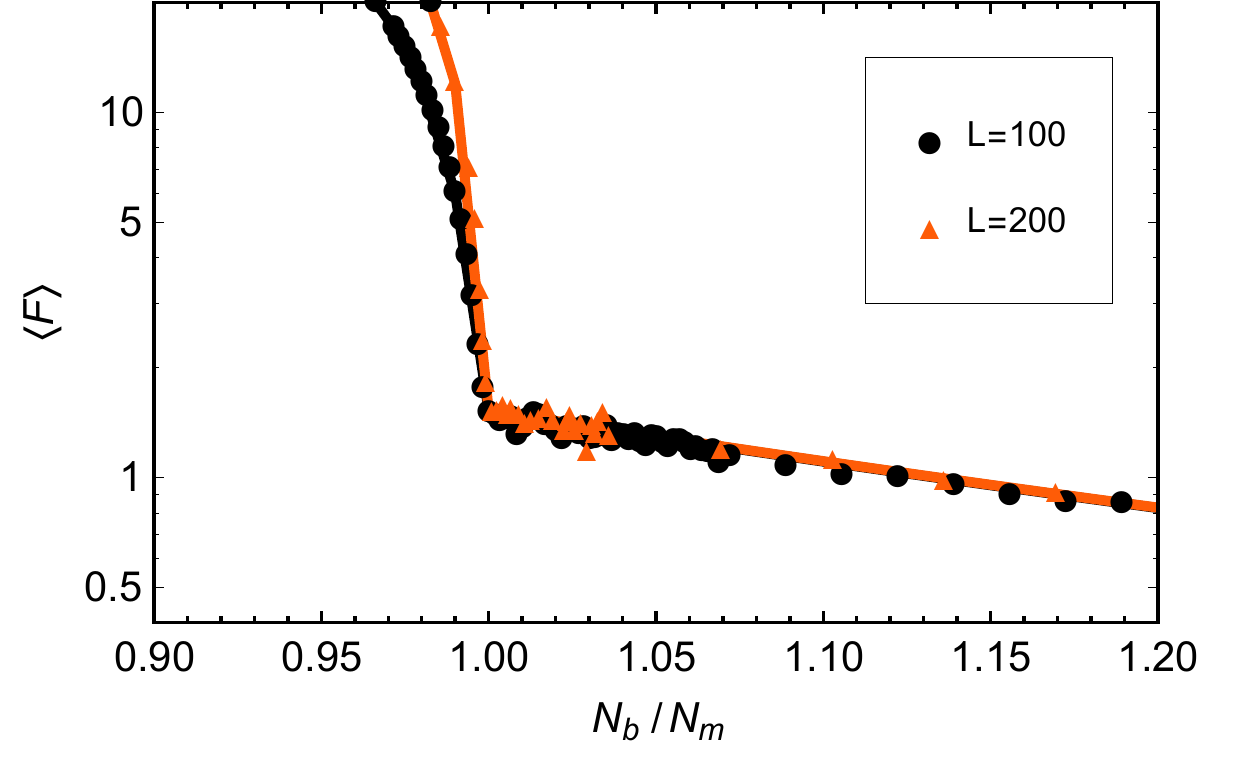}}
\caption{Average number of floppy modes $\avgfloppy$
of (a) generic square and (b) generic kagome lattices as a function
of  $\numbrace/\nummax$ on a semilog plot.  The solid lines show the
theoretical result of Eq.\ (\ref{eq:expf}). There is a
discontinuous change in slope at the Maxwell point $\nummax$. Below
$\nummax$, $\avgfloppy$ is proportional to $\side$, but
above $\nummax$, $\avgfloppy$ no longer scales with $\side$.
\label{fig:floppy modes}}
\end{figure}

From the simulation results, we
find that the threshold for rigidity percolation is at
\begin{align}
\numbraceR^{\textrm{generic}}=\nummax \propto \side,
\end{align}
which occurs much earlier than the rigidity threshold in regular lattices as shown in Eq.~\eqref{EQ:NRRegular}
\footnote{Strictly speaking, as we mentioned in the Introduction, bulk
rigidity occurs at $\numbrace=\nummax-\mathcal{O}(1)$, but for large
lattices this point approaches $\numbrace/\nummax=1$.}. 
As $\numbrace$
increases below $\nummax$, $\probrigid=0$ and $\avgfloppy$
decreases linearly with slope $-1$, following the Maxwell behavior of
Eq.~\eqref{EQ:Maxwell}. At $\numbrace=\nummax$, $\probrigid$ discontinuously jumps to a finite value, and $\avgfloppy$ also exhibits a singularity, due to the sudden appearance of a
rigid bulk. Beyond $\nummax$, $\probrigid$ continues to increase,
while $\avgfloppy$ decreases exponentially to zero. As shown
in Fig.~\ref{fig:pr} and Fig.~\ref{fig:floppy modes}, our data for
$\avgfloppy$ and $\probrigid$ plotted as functions of
$\numbrace/\nummax$ collapse at and above the Maxwell point.
This behavior is explained by our analytical theory in Sec.~\ref{SEC:TheoryGeneric}.  

\begin{figure*}[ht]
\label{genericsnapshots}
\subfigure[$\numbrace-\nummax=-2$]{
\includegraphics[width=0.23\linewidth]{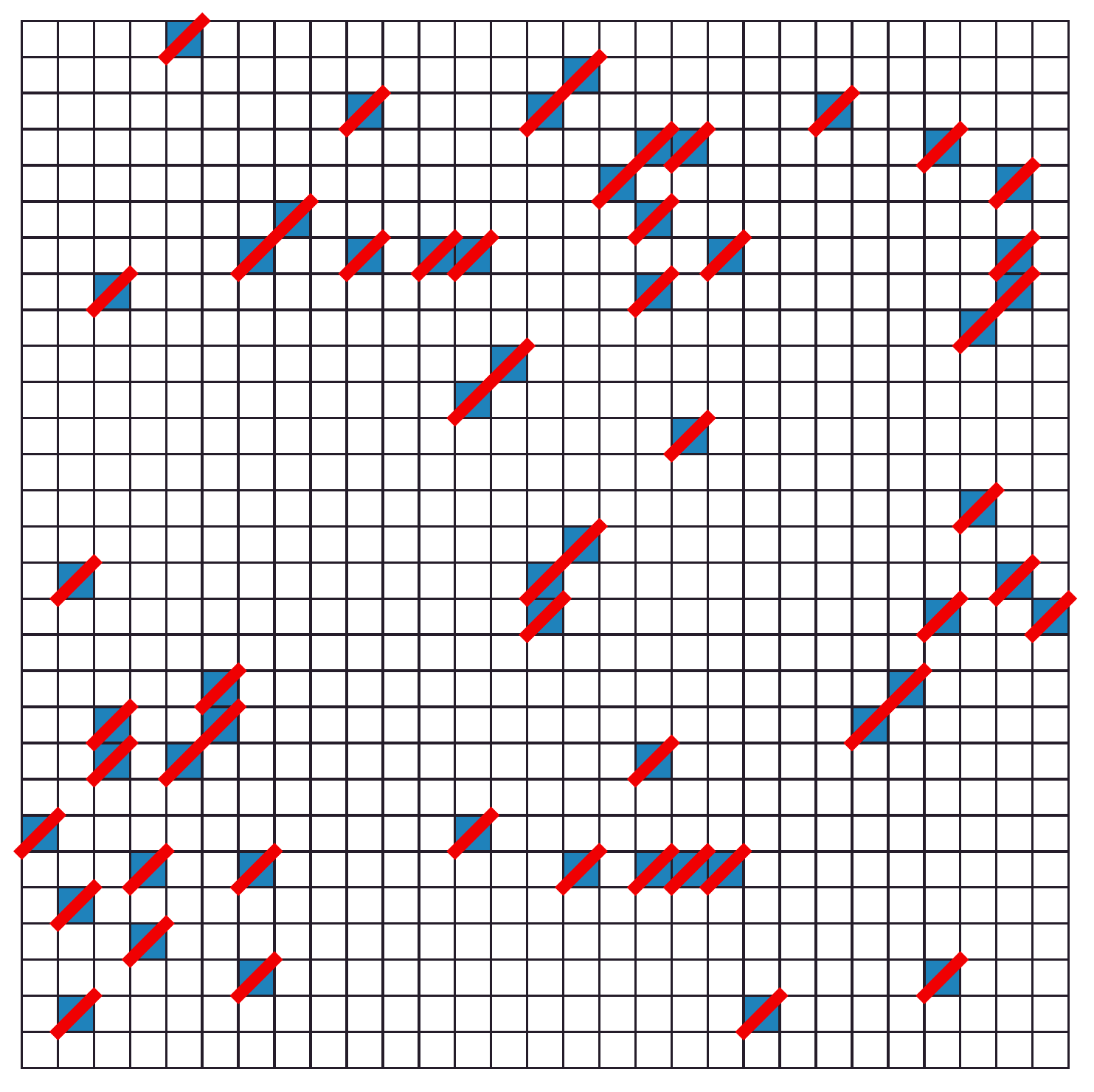}
}
\subfigure[$\numbrace-\nummax=-1$]{
\includegraphics[width=0.23\linewidth]{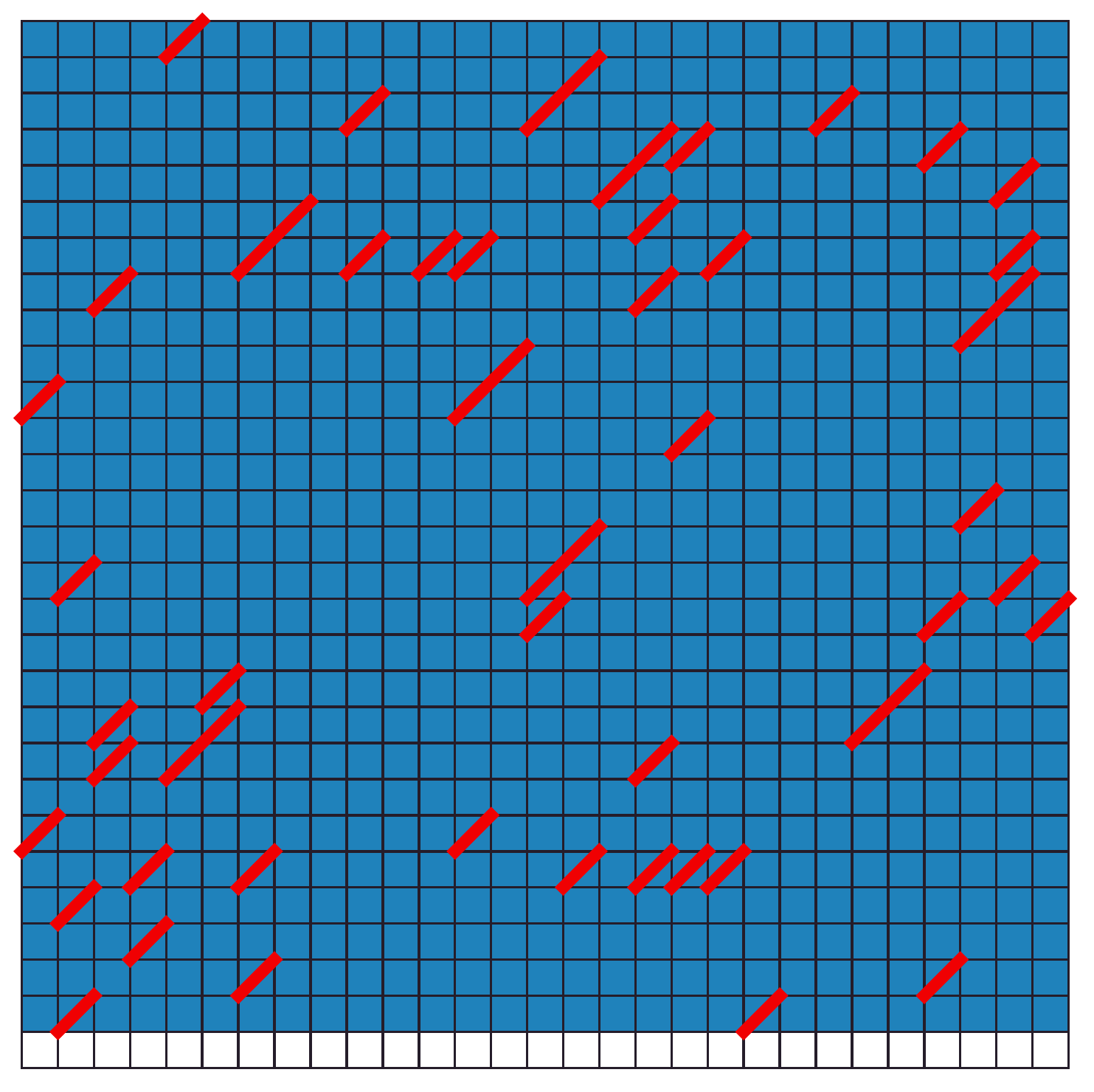}
}
\subfigure[$\numbrace-\nummax=0$]{
\includegraphics[width=0.23\linewidth]{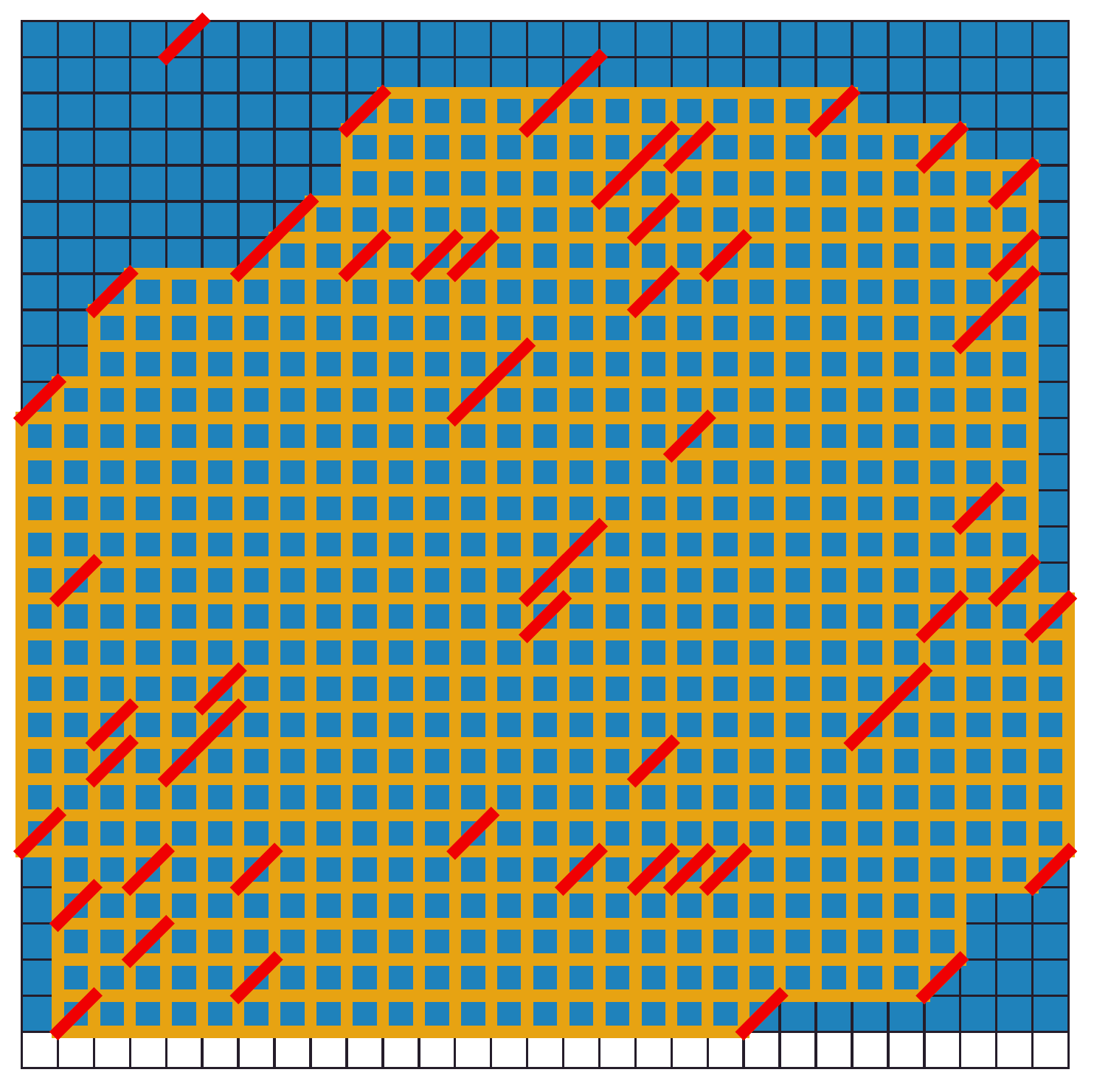}
}
\subfigure[$\numbrace-\nummax=9$]{
\includegraphics[width=0.23\linewidth]{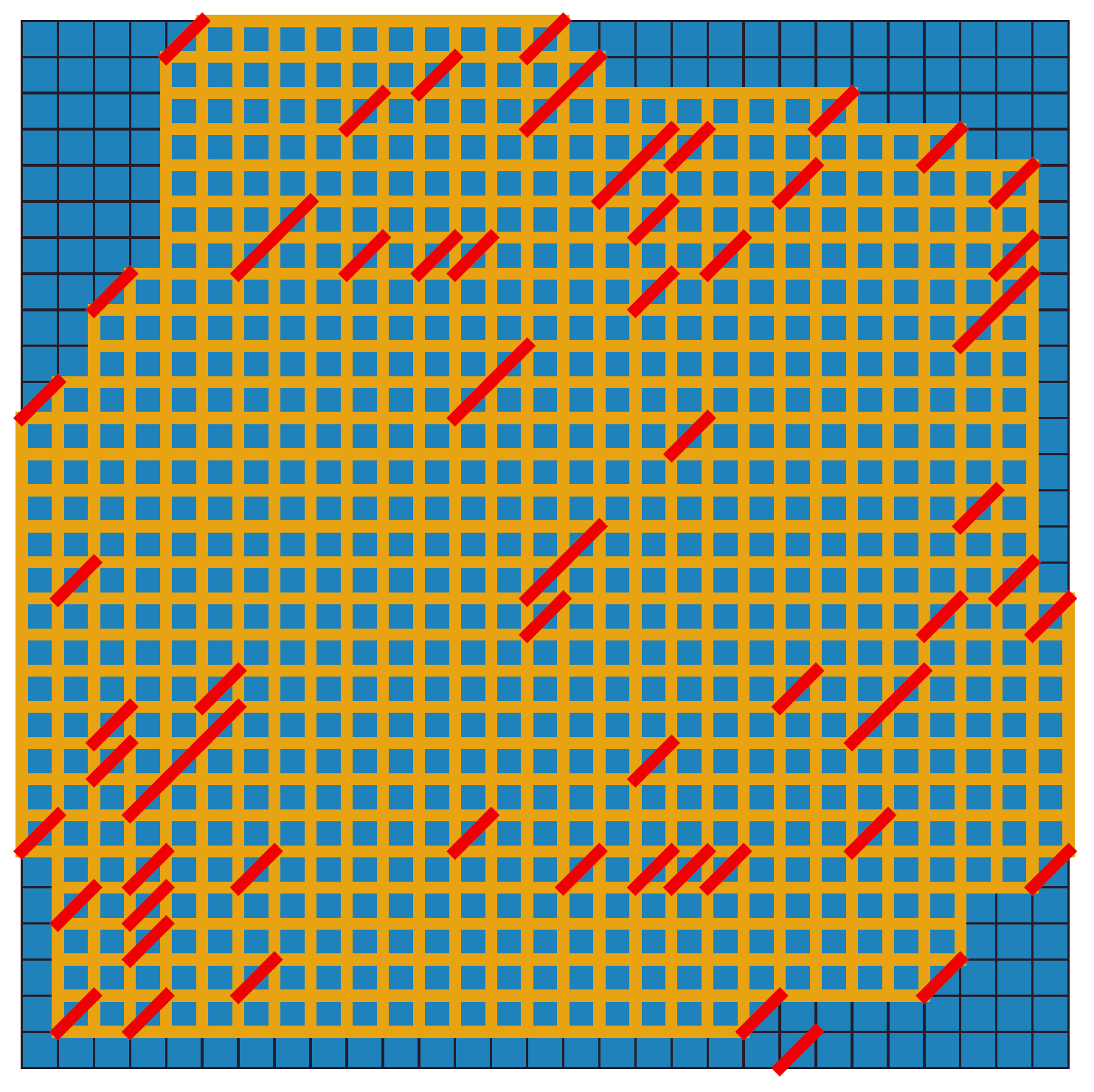}
}
\subfigure[$\numbrace-\nummax=-4$]{
\includegraphics[width=0.23\linewidth]{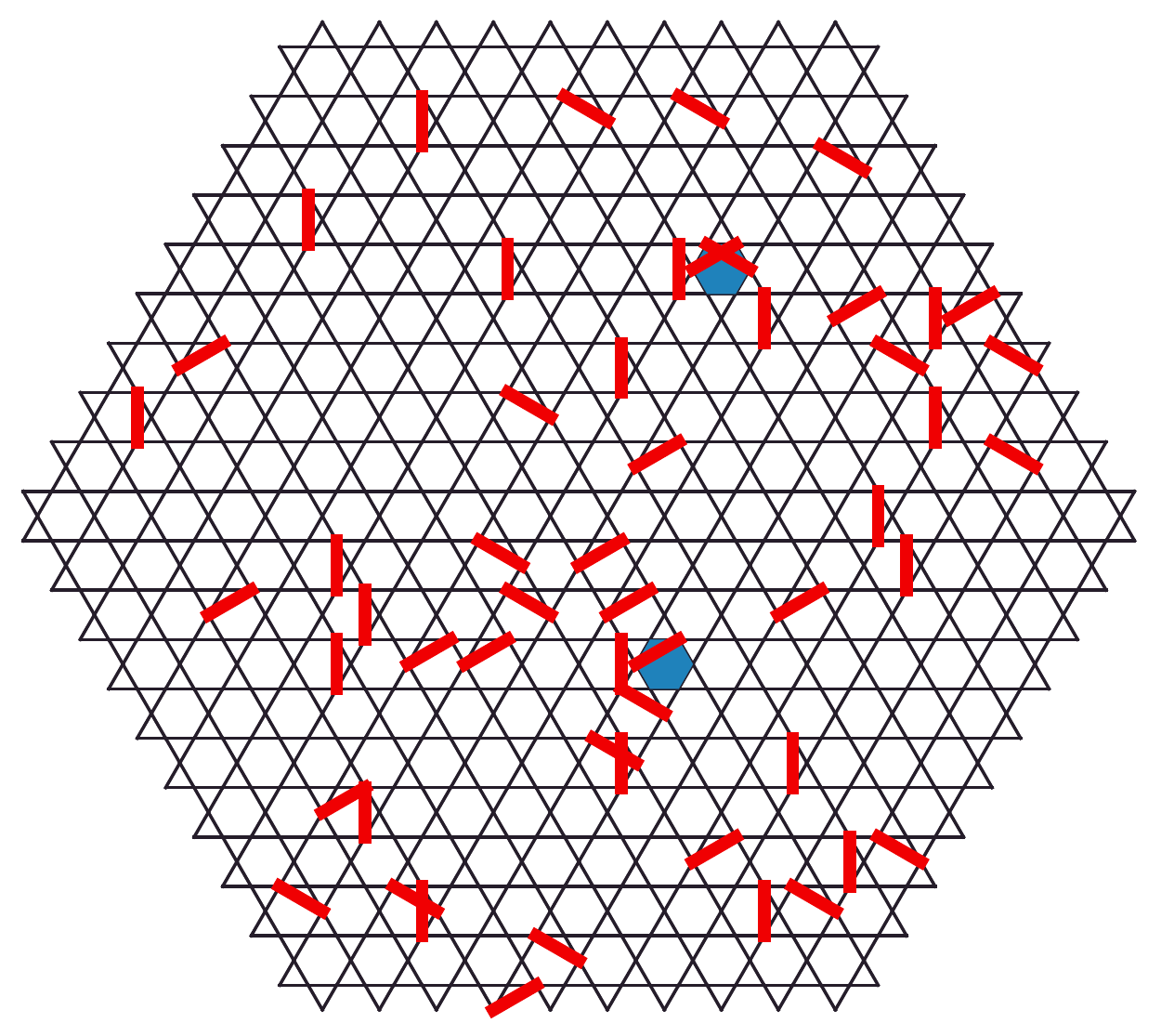}
}
\subfigure[$\numbrace-\nummax=-3$]{
\includegraphics[width=0.23\linewidth]{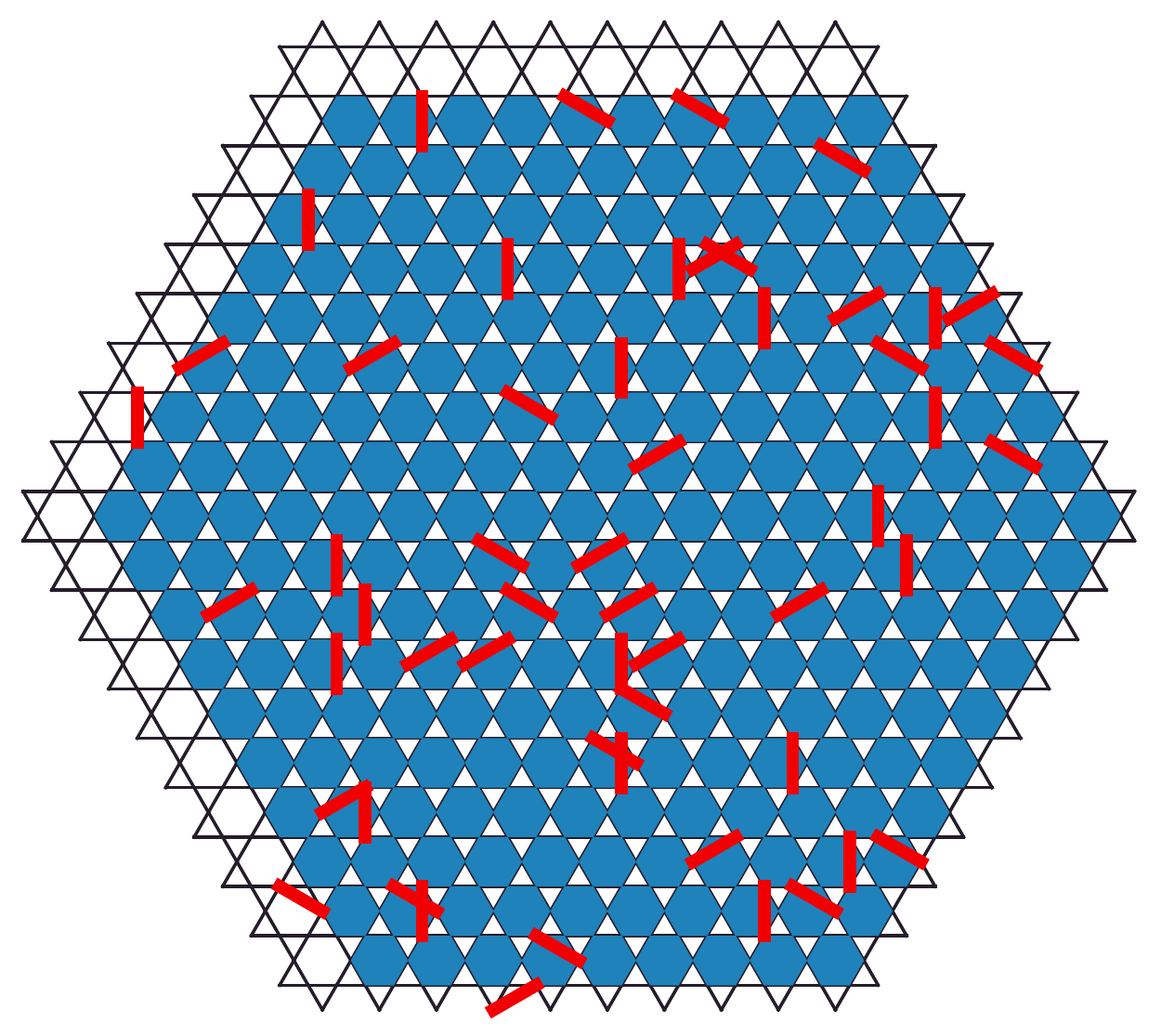}
}
\subfigure[$\numbrace-\nummax=-2$]{
\includegraphics[width=0.23\linewidth]{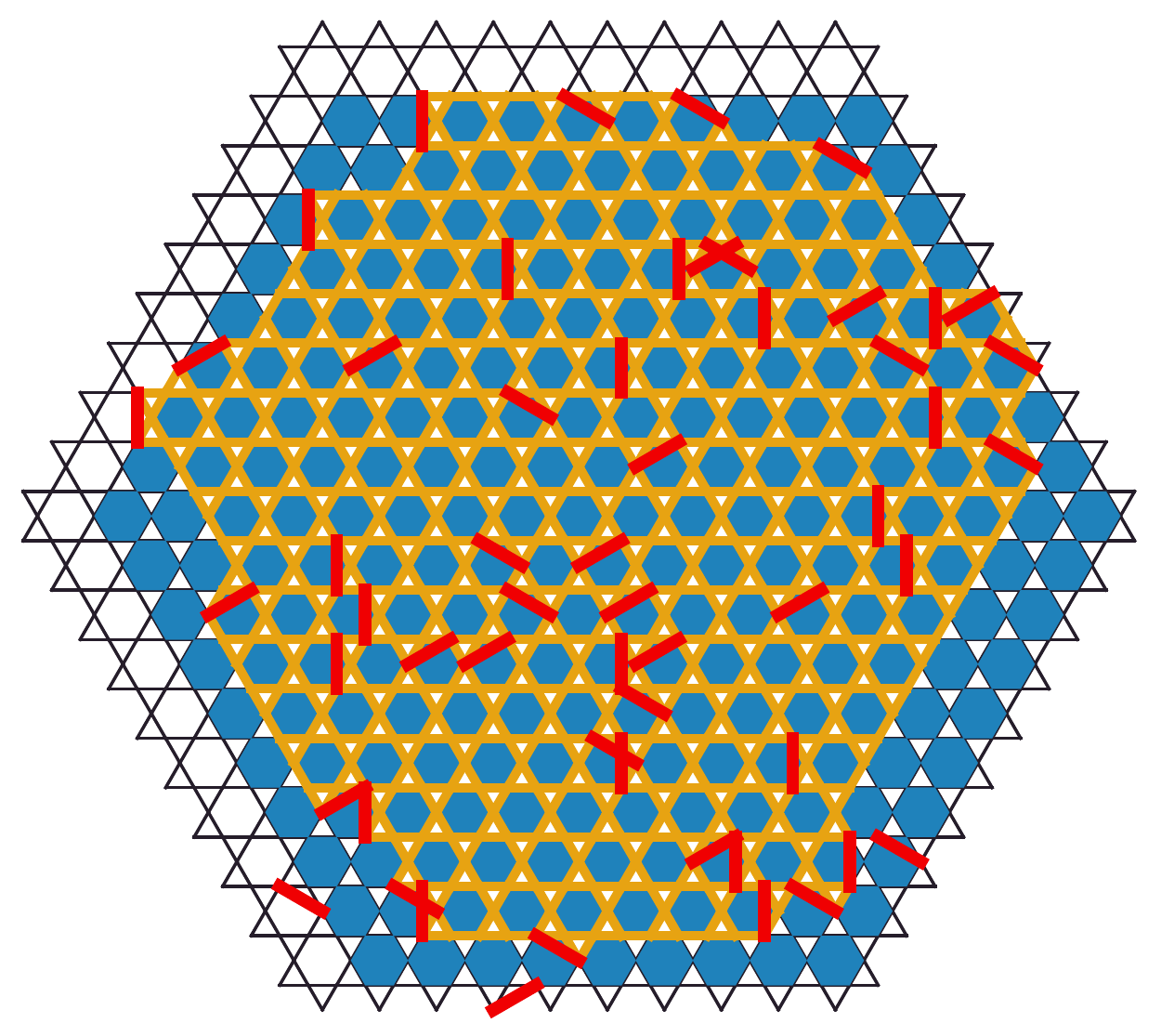}
}
\subfigure[$\numbrace-\nummax=34$]{
\includegraphics[width=0.23\linewidth]{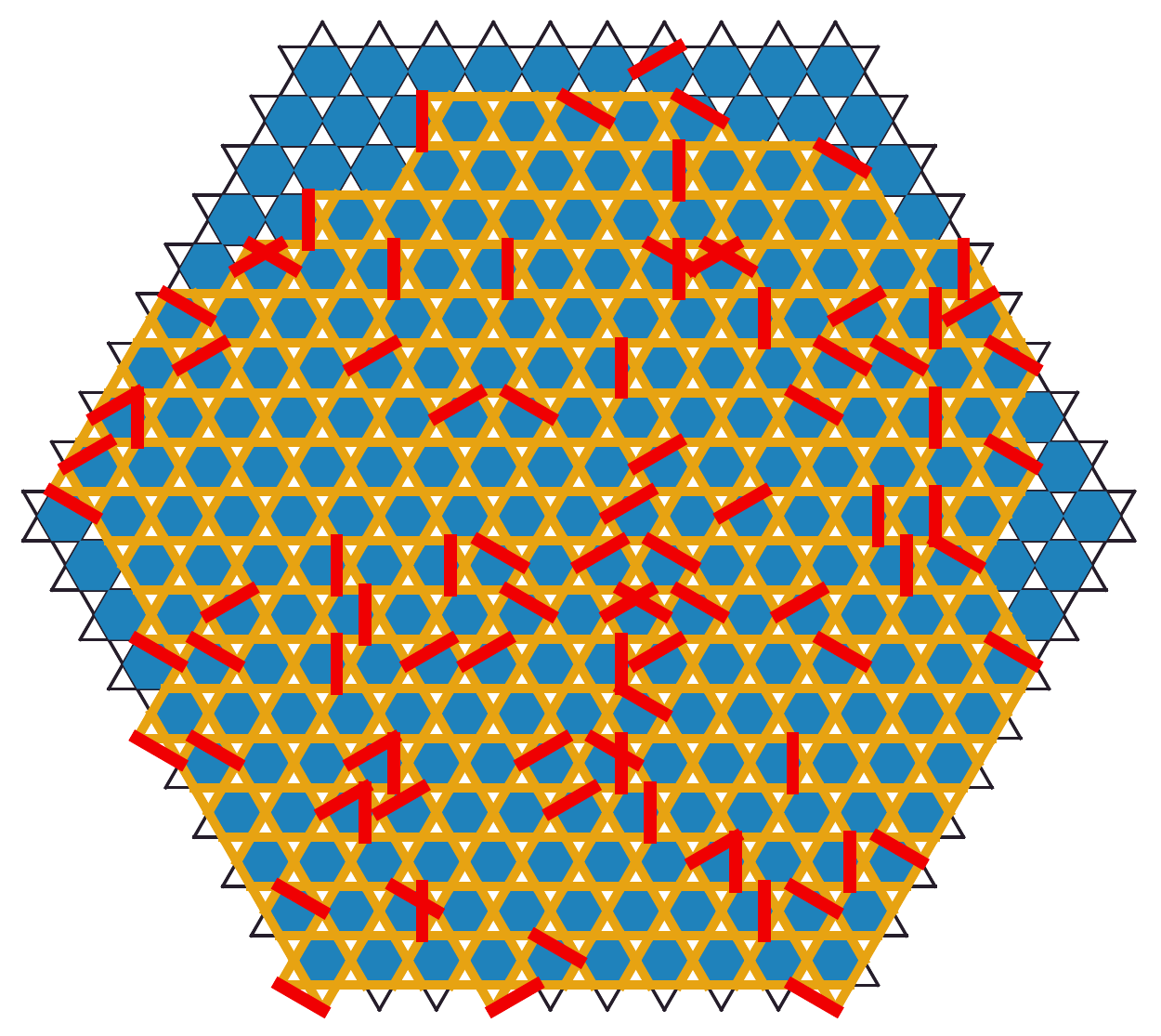}
}
\caption{Snapshots of generic square (a)-(d) and
kagome lattices (e)-(h) with the generic site displacements unpictured for visual clarity.
Randomly-placed braces are shown as red lines, rigid regions as blue areas, and stressed bonds as yellow lines.
In (a) and (e), as braces are added they induce rigidity only locally.
In (b) and (f) a single bond near the Maxwell point has induced
rigidity in the bulk of the system, with at most $\mathcal{O}(\side)$
floppy plaquettes on the edge. As additional braces are added self stresses are generated in the bulk, as shown in (c) and (g). It is only well above the Maxwell point, as shown in (d) and (h) that the floppy modes on the edge are completely eliminated.
\label{fig:rigid_component}}
\end{figure*}

\begin{figure}[h]
\label{regularsnapshots}
\subfigure[$\numbrace-\nummax=-3$]{
\includegraphics[width=0.55\linewidth]{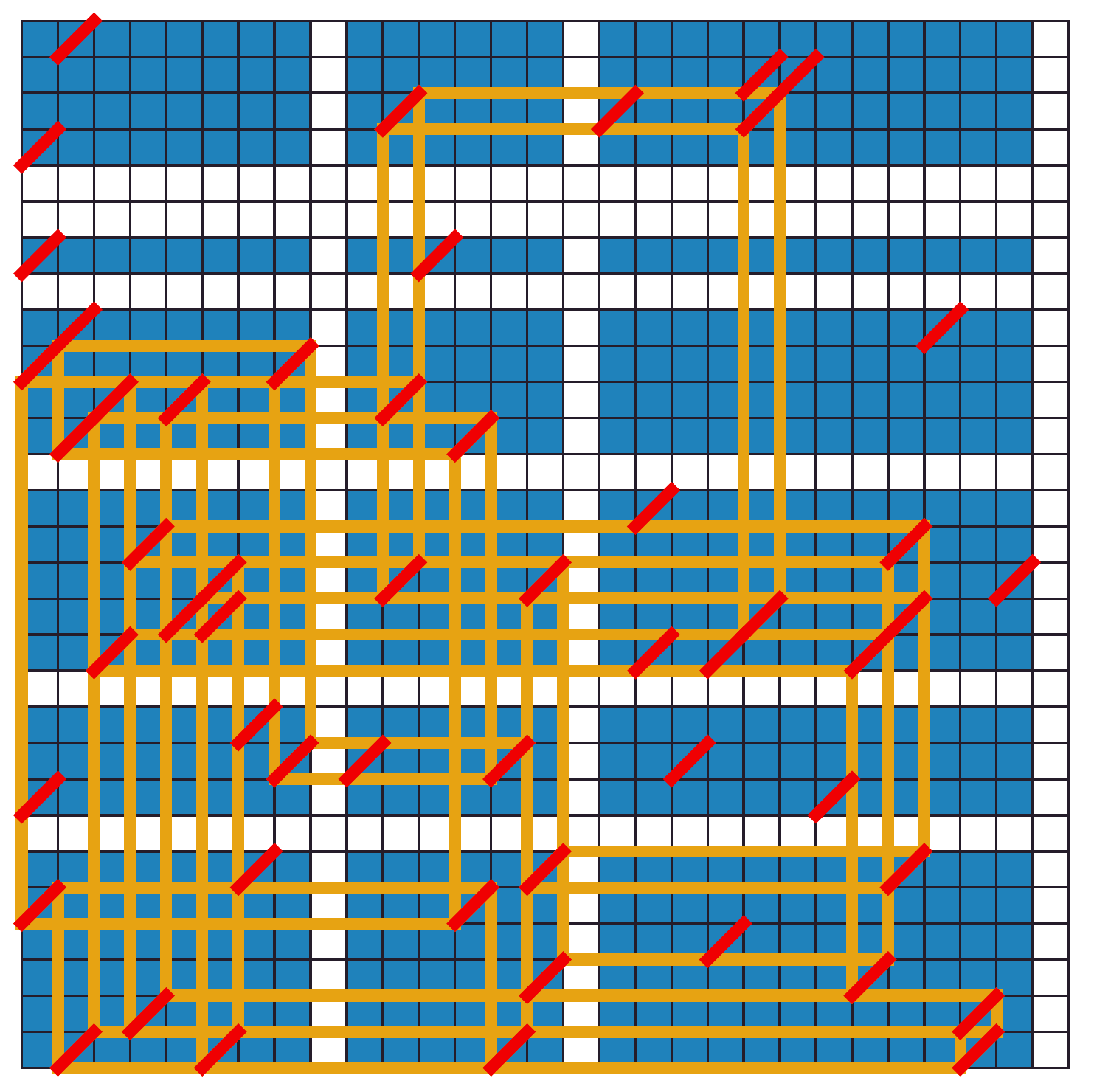}
}
\subfigure[$\numbrace-\nummax=-1$]{
\includegraphics[width=0.65\linewidth]{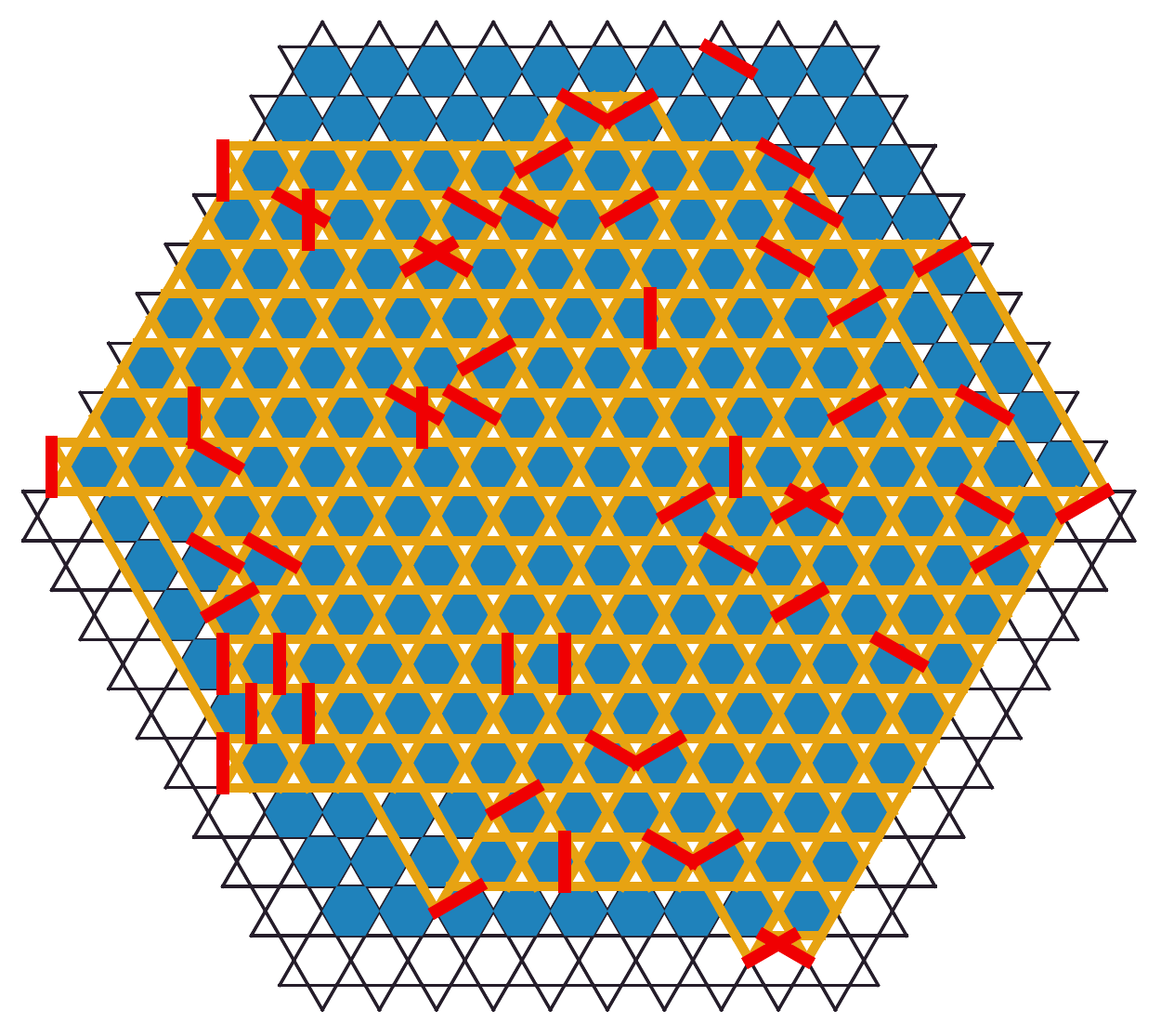}
\label{fig:rigid_regular_kagome}
}
\caption{Snapshots of a regular square lattice (a) and a regular
kagome lattices (b) using the conventions of Fig.\
\ref{fig:rigid_component}. In contrast to the generic lattices,
the regular square lattices (a) feature multiple rigid
components of intermediate size, which are separated by lines of non-rigid regions, instead
of one bulk rigid region. The regular kagome lattice (b) develops
a large rigid component similar to the generic lattices, but
rigidifies much more slowly.
\label{fig:rigid_regular}}
\end{figure}

Fig.~\ref{fig:rigid_component} depicts a sequence of
images showing floppy, rigid, and over-constrained regions in
generic lattices as
$\numbrace$ increases,
illustrating the sudden emergence of a rigid bulk through the addition
of only a single brace (from (a) to (b) in the generic square lattice, and
from (e) to (f) in the generic kagome lattice).  To provide a
comparison, we also include snapshots of rigidity percolation in
regular lattices in Fig.~\ref{fig:rigid_regular}.

\begin{figure}[h]
\label{bulk}
\subfigure[]{
\includegraphics[width=0.8\linewidth]{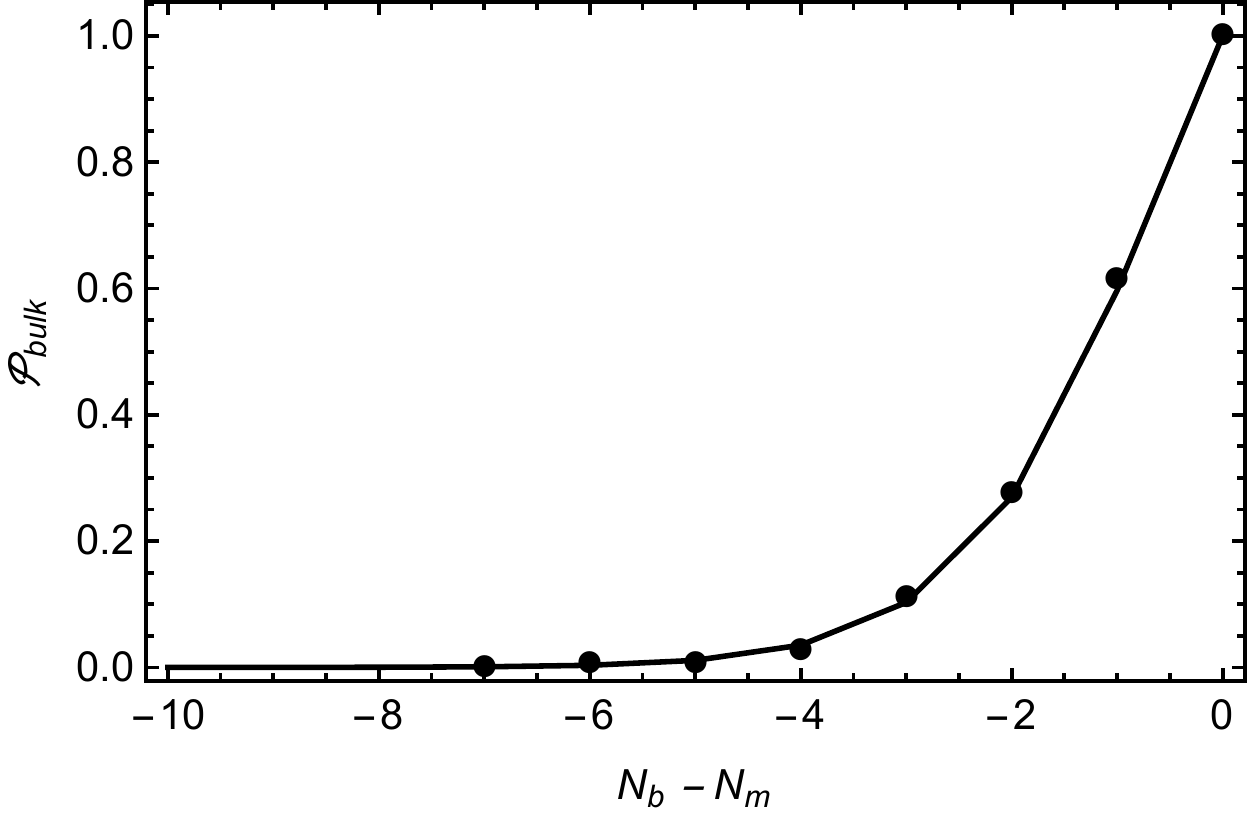}}
\subfigure[]{
\includegraphics[width=0.8\linewidth]{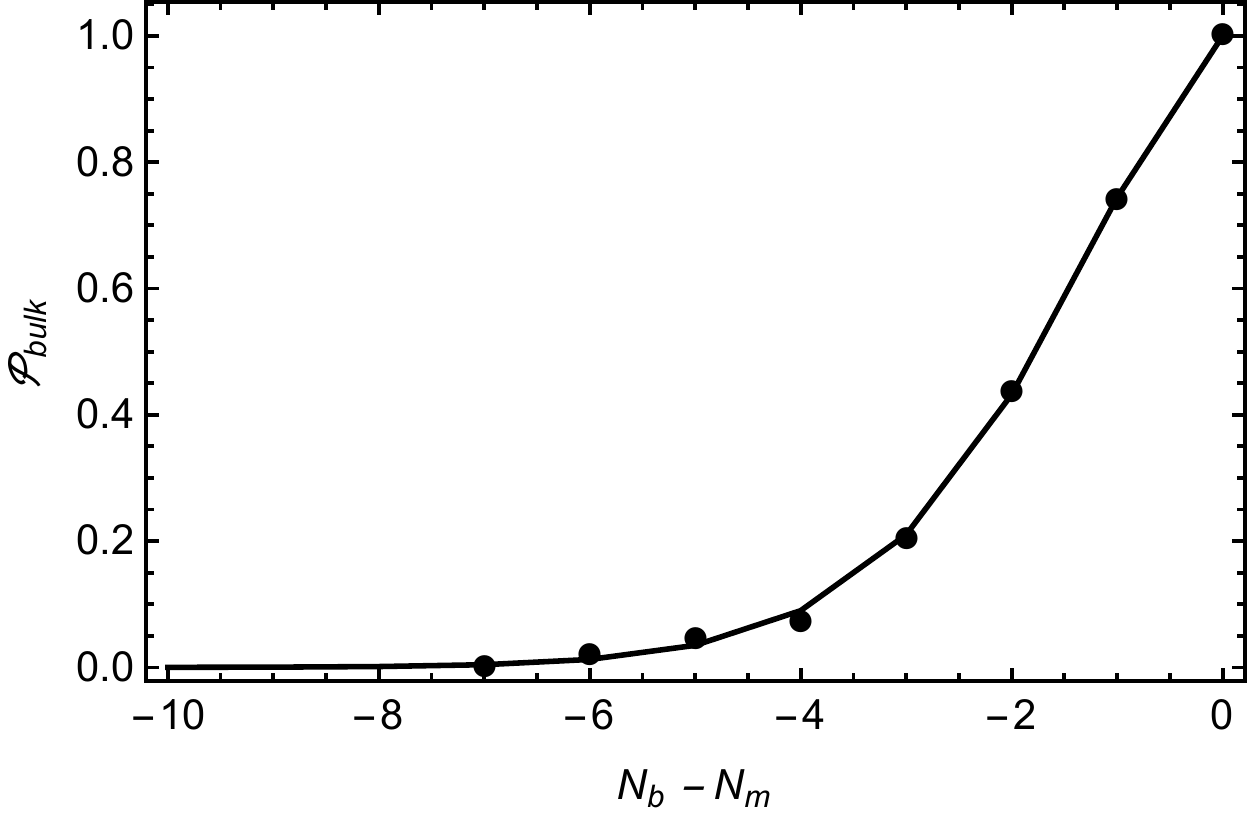}}
\caption{The bulk rigidity probability as a function of
$\numbrace-\nummax$ on (a) generic square lattices with linear size
$\side=200$ and (b) generic kagome lattices with $\side=100$. The solid lines are theoretical results of
Eq.~(\ref{eq:bulkcriterion}) in the large system limit. The dots
are simulation results. Error bars are within the dots.
\label{fig:bulk_rigidity_probability}}
\end{figure}

\section{Theory of rigidity percolation in regular isostatic lattices}\label{SEC:TheoryRegular}
\subsection{Basics of the theory}
In this section we go into detail on rigidity transitions in bracing
percolation on regular lattices.  These lattices are nongeneric in the
sense that there are graphs on the same set of vertices where the
rigidity matrix does not have the maximum possible rank.  Then, if a randomly braced regular lattice is rigid, then the corresponding generic lattice with the same connectivity is also rigid, but the reverse is not always true.

One special property of the braced regular {\em square
lattice} is that its rigidity properties map onto the connectivity
properties of an associated bipartite graph 
\cite{bolkercrapo,Ellenbroek2011}.  
The mapping begins with the observation that an explicit independent
(though not orthogonal) basis for the vector space of zero modes of a
regular square grid of side length $L$ consists of the two global translations as well as
$2L-2$ shears of columns and rows [i.e.\ modes consisting of
translations $(0,1)$ of all vertices with $x\geq j$ (shear of column
$j$) or translations $(1,0)$ of all vertices with $y\geq k$ (shear of
row $k$)].  

Setting aside the translational modes for
now, we assign one vertex of an associated graph to each of the shear
modes (Fig.\ \ref{fig:fmc}).  A brace constraint in the $(j,k)^\textrm{th}$
plaquette couples the shear of column $j$ and the shear of row $k$.  In
the associated graph, we add the edge joining the two corresponding
vertices.  With this construction, the (non-translational) floppy
modes of the braced grid are in correspondence with the connected
components of the associated graph. 

Note that the constructed graph is bipartite, as each potential brace
couples one row shear to one column shear.  In fact, the random
bracing process on a square grid maps to an Erd\H{o}s-R\'enyi process on 
the complete bipartite graph $K_{L-1,L-1}$~\cite{Ellenbroek2011}.  

For regular kagome lattices, there is no such mapping.
However, as explained in Appendix A, and exploited in our simulations
described in Sec.\ \ref{SEC:Simulation},
we can still find a line-localized basis of independent zero modes where
the displacements for each of these modes are localized onto particles
on separate straight lines in the kagome lattice. This basis has the advantage
 that each brace couples only four modes together, which allows
us to analyze the rigidity of the regular kagome via the interaction
of independent braces on the line modes.

As mentioned in Sec.~\ref{Sec:SimuRegu}, the two observables
$\probrigid$ and $\avgfloppy$ undergo two {\em
distinct} transitions in braced regular lattices.  Following the notion in Ref.~\cite{Ellenbroek2011}, the rigidity percolation transition can be defined as the point where $\probrigid$ exceeds $1/2$.
As we discuss below, this occurs in the regular kagome lattice when the number of added braces $\numbraceR \sim \side \ln \side$, same as the result found
for regular square lattices~ \cite{Ellenbroek2011}, and explains our observation from simulation in Sec.~\ref{Sec:SimuRegu}.

The other transition occurs at a lower $\numbrace$ where the number of floppy modes has a singularity in the large $L$ limit when the number of
added braces $\numbraceR \sim \side$. The nature of this singularity
differs in the regular square and regular kagome lattices, so we
describe them in turn here. 

Previous work has suggested that the number of floppy modes is the
analog of a ``free energy'' in rigidity percolation systems
\cite{duxbury}. Indeed, in the aforementioned
mapping between the regular square lattice with braces and bipartite graphs, the 
number of floppy modes in this rigidity percolation problem maps precisely to the number of connected
components in a connectivity percolation problem.  
Thus, in the regular square lattice there is a singularity in $\avgfloppy$
corresponding to the
formation of a giant component in the associated bipartite graph, and it occurs at
\begin{align}
	\numbraceG^{\textrm{regular square}}=\nummax/2 ,
\end{align}
well before the Maxwell point. 
We will show that this occurs via a second-order mean field transition, and in fact the
singularity is actually a discontinuity in the {\em third} derivative
of $\avgfloppy$, which is not visible in Fig.\
\ref{fig:regular_floppy_plots}(a).

In the regular kagome lattice, the situation is rather different.  There is a visible 
kink in $\avgfloppy/L$ at the Maxwell point (Fig.\ \ref{fig:regular_floppy_plots}(b)),
\begin{align}
	\numbraceG^{\textrm{regular kagome}}=\nummax .
\end{align}
This indicates a first-order transition there, which is associated to
the formation of a single giant rigid cluster.

The reason that there are two distinct transitions in the regular
lattices is because sets of braces can form redundancies
fairly easily. The coupling of a large number of floppy modes
together is not sufficient to {\em completely} rigidify the
system.  In the regular square lattice, this coupling does not even
create a single large rigid component, though it does 
for the regular kagome lattice.
Despite the fact that most of the floppy modes become coupled
together, many floppy modes remain ``isolated''---that is, decoupled from all 
other modes---and added braces tend to create redundancies rather than
remove degrees of freedom.  In the bipartite graph representation of
the regular square lattice, these floppy modes correspond indeed to isolated
vertices, and adding enough braces to the system to couple them all to
the giant floppy mode yields a coupon-collector problem \cite{feller}.
The rigidity transition in the regular kagome lattice proceeds
through a similar, but more complicated process. Thus in both cases, there
is a separate transition to rigidity of the system which occurs much 
later: when $p\sim \ln L/L$. 

Note that in both transitions there are system sizes which diverge
with $1/p$ (to lowest order) as the probability goes to zero, below which the system is very
floppy, and above which the system is rigid.

\subsection{Probability of rigidity in regular lattices}
We first describe the situation for the regular square lattice,
giving a heuristic derivation of the results of \cite{Ellenbroek2011}.  Next, using those ideas, we
conjecture a generalization for the regular kagome lattice which
conforms closely to our simulation results with no free parameters.
The probability of rigidity in a regular square lattice is simply the
probability that there is a single connected component of the
Erd\H{o}s-R\'enyi model.
This is asymptotically equal for large $L$ to the probability of having at least one
brace in every row and column of plaquettes \cite{palasti}.  Though this is 
a necessary rather than sufficient condition for rigidity~\cite{Ellenbroek2011}, the probability of having a nonrigid configuration satisfying this condition goes to zero as the system size gets large.

Assuming this, and neglecting the slight dependence between the events of having a brace in a row and having a brace in a column, the probability that the configuration is rigid is
the product over all rows and all columns of the probability that
there is at least one brace in that row or column:
\begin{align}
\probrigid(p)
&=\prod_{i\text{ rows, columns}}[1-(1-p)^{\side-1}]\nonumber\\
&=[1-(1-p)^{\side-1}]^{2L-2}\nonumber\\
&\approx e^{-2\side e^{-p\side}}.\label{eq:regsqrigprob}
\end{align}

The underlying probabilistic process resembles that of the
coupon-collector problem \cite{feller}, where supposing there are $n$ distinct
and equally likely coupon types, one asks how many coupons must be received
before all $n$ types have been seen at least once.
This heuristic is in perfect agreement with the results of Ref.~\cite{Ellenbroek2011} asymptotically as $\side\rightarrow\infty$; we
compare to numerical simulations in
Fig.\ \ref{fig:regular_prob_plots}(a).  We define the threshold
probability $\rigidthreshold$
 via
$\probrigid(\rigidthreshold)=\frac{1}{2}$. In the limit $L\rightarrow\infty$ we
have 
\begin{align}
\rigidthreshold^{\textrm{regular square}}=&\ln \side/\side+\ln(2/\ln 2), 
\end{align}
and clearly the right scaling
variable for this transition is $p \side/\ln \side$, meaning that $\probrigid$ changes appreciably when
$p-\rigidthreshold^{\textrm{regular square}}\sim \mathcal{O}(\ln \side/\side)$.

We now postulate that, asymptotically, the regular kagome lattice in a
hexagon becomes rigid precisely when every line meets at least one brace.
One complication in repeating the above calculation is that the number
of possible braces per ``line'' in the kagome lattice hexagon is
not uniform, as the lines are of different length with our boundary
conditions.  

Thus let us first count the number of possible braces per line. There are three possible
directions of lines: they lie at angles $0,2\pi/3,4\pi/3$ relative to
the $+x$ axis, and we can further divide the set of lines with a
given slope into two families which lie on opposite sides of the line
cutting the hexagonal domain in half. These two families are related
to each other by a reflection symmetry across that line.

We find that within one of these families, the line on the
boundary admits $l_1=3\side+1$ possible
braces, and the other $\side-1$ interior lines have length $l_m=4\side+4m-3$
for $m=2$ to $\side$. We multiply these counts by 6 because of the
aforementioned 6-fold symmetry.

Proceeding as we did for the square lattice:
\begin{align}
\probrigid(p)&=\prod_{i\text{ lines}}[1-(1-p)^{l_i}]\nonumber\\
&\approx[1-e^{-p(3\side+1)}]^6\prod_{m=2}^\side[1-e^{-p(4\side+4m-3)}]^6\nonumber\\
&=[1-e^{-p(3\side+1)}]^6e^{6\sum_{m=2}^\side(1-e^{-p(4\side+4m-3)})}\nonumber\\
&\approx[1-e^{-3p\side}]^6e^{\frac{3}{2p}\left[\operatorname{Li}_2(e^{-8p\side})-\operatorname{Li}_2(e^{-4p\side})\right]}.\label{eq:regkagrigprob}
\end{align}
In the last expression,
$\operatorname{Li}_2(z)\equiv\sum_{k=1}^\infty\frac{z^k}{k^2}$ is the
dilogarithm function, arising from approximating the sum as an integral.  
This expression compares well with the results from numerical simulations, depicted
in Fig.\ \ref{fig:regular_prob_plots}(b), which a posteriori justifies
our assumption above.  

We now evaluate $\rigidthreshold^{\textrm{regular kagome}}$ in the limit $\side\rightarrow\infty$:
\begin{align}
\frac{1}{2}&\approx[1-e^{-3\rigidthreshold
\side}]^6e^{\frac{3}{2\rigidthreshold}\left[\operatorname{Li}_2(e^{-8\rigidthreshold
\side})-\operatorname{Li}_2(e^{-4\rigidthreshold \side})\right]}\nonumber\\
-\ln 2&\approx
\frac{3}{2\rigidthreshold}\left[-e^{-4\rigidthreshold \side}\right].\label{eq:pcdef}
\end{align}
After taking the logarithm, we keep only the lowest powers of
$e^{-\rigidthreshold \side}$ in each
factor.  Physically, this corresponds to neglecting boundary effects and
the variation in the line lengths and noticing that in the large $L$ limit,
the rigidity threshold is approached once the longest lines in the
hexagon are coupled to the bulk:
\begin{align}
\ln 2&\approx \frac{3 }{2 \rigidthreshold}e^{-4 \side \rigidthreshold}\nonumber\\
\rigidthreshold&=\frac{1}{4 \side}W\left(\frac{6\side}{\ln 2}\right).
\end{align}
The function $W(\cdot)$ is the Lambert W function, defined to be the
solution of $x=W(x)e^{W(x)}$. We find that the approximation above
matches the solution in Eq.\ (\ref{eq:pcdef}) to high accuracy only
when $\side>10^{8}$.
The asymptotic expansion for $W(x)$ begins $W(x)\sim\ln
x-\ln\ln x$,  thus in the limit
$\side\rightarrow\infty$, we find that:
\begin{align}
\rigidthreshold^{\textrm{regular kagome}}\sim&\frac{\ln \side}{4\side}. 
\end{align}
The corrections to this do not go to zero but rather grow more slowly
in $\side$ than $\ln \side/\side$.
Regardless, this shows that the rigidity transition in the regular
kagome hexagon resembles that of the regular
square grid in that it occurs roughly after adding $\mathcal{O}(\side\ln
\side)$ braces.

Our calculation thus show that $\probrigid(p)$ approaches a discontinuous jump as $\side\to\infty$, which signals a first-order transition.  On the other hand, one can extract a diverging length $\xi^{\textrm{regular}}\sim 1/p$ near the transition, signaling a second order transition.  Therefore this model relates to a group of interesting systems that exhibit such mixed nature~\cite{Berthier2011,Mezard1999,Liu2010,Schwarz2006,aizenmanlebowitz,holroyd,kineticallyconstrainedreview}.

\subsection{Number of floppy modes in regular lattices}
The picture that follows from our assumptions and the
calculation above is that at large $p\side$, the floppy modes
of the regular lattice systems consist of one large coupled floppy
mode and many isolated modes. This idea also allows us to calculate the
number of floppy modes as the system approaches rigidity.

In the regular square grid, we expect that for large $p\side$, the
average number of floppy modes $\avgfloppy$ is the sum over all
lines of the probability that the corresponding mode is not coupled to
any others, i.e.\ that the line meets no braces. As each of the
$2\side$ lines has length $\side$, this predicts that 
\begin{align}
\avgfloppy &\rightarrow 2\side(1-p)^L\nonumber\\ 
&\approx 2\side e^{-p\side}.\label{eq:asymptoticregularsquarefloppy}
\end{align}

In Appendix B, we exploit the mapping
to the bipartite Erd\H{o}s-R\'enyi model to
derive the following expression for $\avgfloppy/\side$
[Eq.\ (\ref{eq:solution})] that is valid for all $p\side$:
\begin{align}
\frac{\avgfloppy}{\side}=&2(1-s_*(p\side))\left(1-\frac{p\side}{2}(1-s_*(p\side))\right),\label{eq:regularsquarefloppy}
\end{align}
with $s_*(p\side)$ defined to be the stable solution of
$1-s_*=e^{-p\side s_*}$~\cite{engel}.  In particular, this
reduces to Eq.\ (\ref{eq:asymptoticregularsquarefloppy}) in the limit
$p\side\rightarrow\infty$. Fig.\
\ref{fig:regular_floppy_plots}(a) compares Eqs.\ (\ref{eq:asymptoticregularsquarefloppy}) and
(\ref{eq:regularsquarefloppy}) to the results from numerical
simulations.

Note that $s_*$, which is the probability that a given line mode is
coupled to the ``giant component'' floppy mode (analogous to the
magnetization in the Potts model~\cite{wupercolation}, see also
Appendix B) has a kink at $p\side=1$ (see Fig.\ \ref{fig:sstar}). For
$p\side<1$, $s_*$ is identically
0, but then begins to grow linearly for $pL>1$.  This value of $pL$ corresponds
to the addition of only {\em half} of the braces required to get to
$\nummax$.  The discontinuity in $s_*$ leads to a discontinuity in the
third derivative of $\avgfloppy/\side$, which is not visible
in the plot of Fig.\ \ref{fig:regular_floppy_plots}(a). This
singularity corresponds to the formation of a giant component in the
bipartite graph from the mapping. In rigidity terms, it is a
transition where one floppy mode couples a large number of the row and
column shear modes together, i.e. the formation of a ``giant floppy
mode''.  From the mapping (or direct calculation from Eq.\
(\ref{eq:regularsquarefloppy})), the critical exponents for this giant
floppy mode transition are the same as for mean-field 
percolation~\cite{StaufferAha1994}.

For the regular kagome lattice, we observe
a more dramatic transition at $\numbrace=\nummax$. Fig.\
\ref{fig:regular_floppy_plots}(b) shows a kink in
$\avgfloppy/\side$ there, implying a discontinuity in the first
derivative. Such a discontinuity can be interpreted in the following
way. 
Eq.\ (\ref{eq:calladine}) implies that this kink in $\numfloppy$ comes from a kink in the number of self stresses. Since self stresses occur when a bond is placed in a rigid region, this discontinuity implies a discontinuous jump in the density of rigid regions in the system~\cite{moukarzelfield}.
  Based on our numerical experiments, we
observe that this corresponds to the formation of a \emph{single} large rigid
cluster in the bulk, as shown in Fig.~\ref{fig:rigid_regular_kagome}. 

Analytically, we have only been able to compute an asymptotic result for
$\avgfloppy/\side$ for large $p\side$. As in our asymptotic form in Eq.~(\ref{eq:asymptoticregularsquarefloppy}), we expect
that at large $p\side$ the expected number of floppy modes $\avgfloppy$ is 
the sum over all lines of the probabilities that each line meets no
braces.  We worked out these probabilities in the previous section
(though we computed the {\em product} of the {\em
complementary} probabilities in Eq.\ (\ref{eq:regkagrigprob})). For
large $p\side$, we have 
\begin{align}
\avgfloppy&=\sum_{i\in \text{lines}}(1-p)^{l_i}\nonumber\\ 
&\approx 6e^{-p\side} + 6\sum_{m=2}^\side e^{-p(4\side+4m-3)}\nonumber\\
&= 6e^{-4p\side} + 6e^{-4p(\side-1)}\frac{1-e^{-4p(\side-1)}}{1-e^{-8p}}\nonumber\\
&\approx
3\side e^{-4p\side}\frac{1-e^{-4p\side}}{2p\side}.\label{eq:asymptoticregularkagomefloppy}
\end{align} 
We find that this matches well the numerical
results as soon as $\numbrace>\nummax$, depicted in Fig.\
\ref{fig:regular_floppy_plots}(b). 

One important fact which is apparent in the above analysis is that in
both the regular square and regular kagome systems, the number of
floppy modes remains ``extensive'', i.e. scales with the linear system
size $\side$, for all parameter values.  We shall see in the next
sections that this is not true in the generic systems with rigid bulks.

\section{Theory of rigidity percolation on generic isostatic
lattices}\label{SEC:TheoryGeneric}
\subsection{Formation of rigid regions}
In this section we develop an analytical theory to predict where and how
rigidity develops as braces are randomly added to the system. We focus
on the generic square lattice, with results that are readily extended
to the kagome lattice. As we will see, added braces pose independent
constraints on the system's floppy modes until close to the Maxwell
point, when a single brace makes the entire bulk of the system rigid (see Fig.~\ref{fig:rigid_component}). Once the bulk is rigid, only the edges can contain floppy modes, and these edge modes may persist well above the Maxwell point.

\begin{figure}
\subfigure[ Regular lattice, neighboring bonds]{
\includegraphics[width=0.45\linewidth]{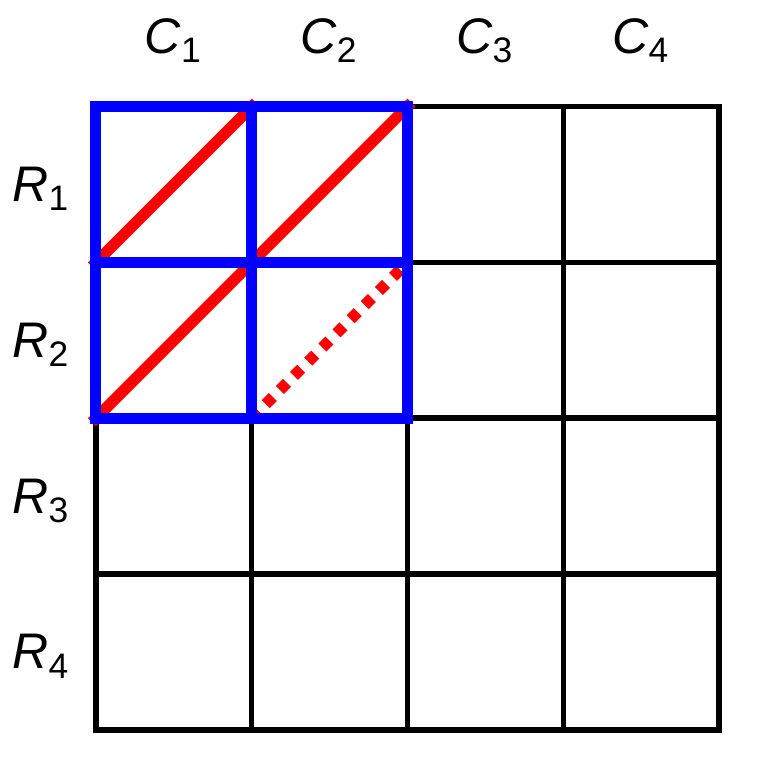}
}
\subfigure[ Generic lattice, neighboring bonds]{
\includegraphics[width=0.45\linewidth]{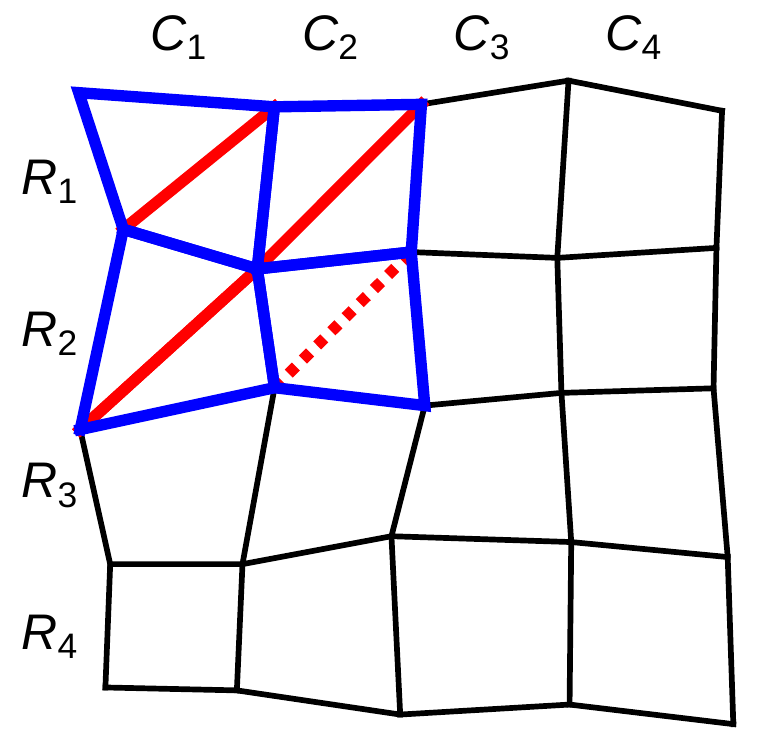}
}
\subfigure[ Regular lattice, distant bonds]{
\includegraphics[width=0.45\linewidth]{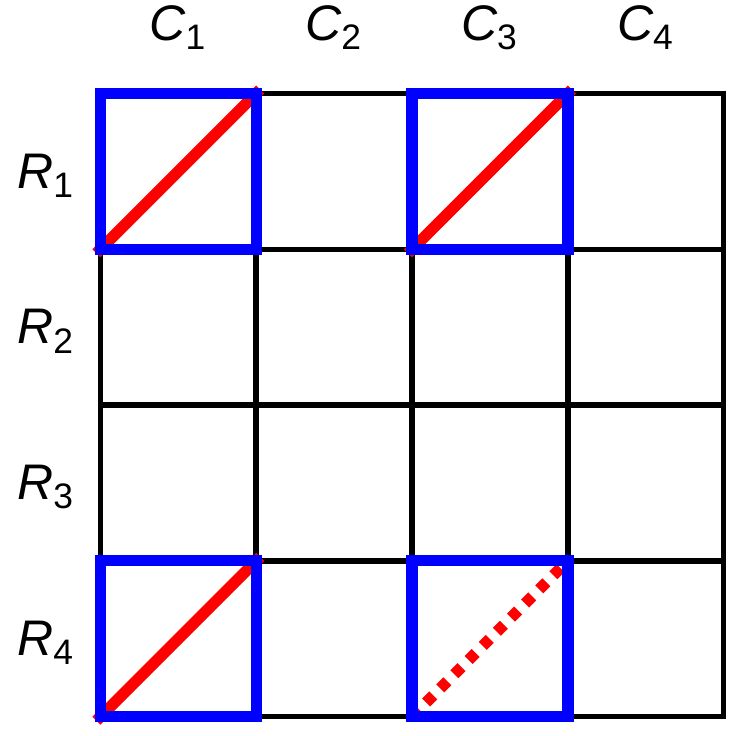}
}
\subfigure[ Generic lattice, distant bonds]{
\includegraphics[width=0.45\linewidth]{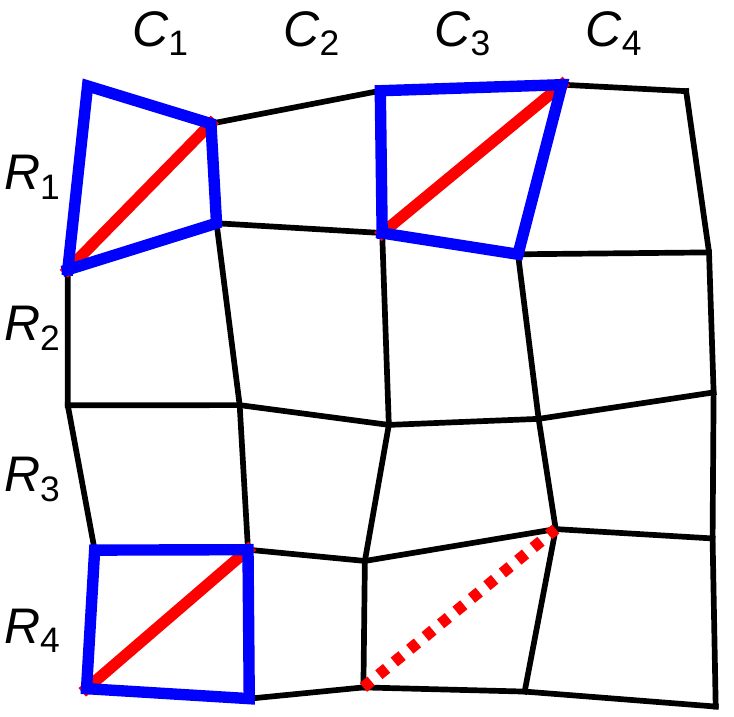}
}
\caption{
Regular and generic lattices differ dramatically in how individual
plaquettes become rigid. As shown in (a) and (b), in either type of
lattice three braced plaquettes render the fourth plaquette that
shares a vertex with them rigid. Because the floppy modes of the
regular lattice shear whole columns or rows, three braced plaquettes
can also render a fourth distant plaquette rigid, meaning that an
additional brace placed there would generate a self stress. In (c),
the plaquette with the dashed line is rigid because shearing it would
require rotating the plaquettes at $(R_4,C_1)$ and $(R_1,C_3)$ to
different angles, which would then shear the braced plaquette at
$(R_1,C_1)$. In contrast, the generic mixing of floppy modes in (d)
means that the trio of braced plaquettes do not render $(R_4,C_3)$
rigid, and so that when a brace is placed there it eliminates a floppy mode, rather than generating a self stress as in the equivalent regular lattice.
}
\label{fig:regvsgen}
\end{figure}

Rigidity percolation on the generic square lattice differs from
that of the regular square lattices in an important way.  
It is worth noting that the bipartite graph mapping for the regular
square lattice does not preserve
any information of distances between rows and columns.  For example,
if three braces join row $\mathcal{R}_i$ with column
$\mathcal{C}_k$, row $\mathcal{R}_i$ with column $\mathcal{C}_l$, and
$\mathcal{R}_j$ with column $\mathcal{C}_k$, then
$\mathcal{R}_i,\mathcal{R}_j,\mathcal{C}_k,\mathcal{C}_l$ already
belong to the same rigid cluster, and the addition of a brace at
the plaquette of $\mathcal{C}_l$ and $\mathcal{R}_j$ must be
redundant, no matter how far the distance is between $\mathcal{R}_i$
and $\mathcal{R}_j$, and $\mathcal{C}_k$, and $\mathcal{C}_l$.  For a
generic square lattice, in contrast, such a fourth brace is only
redundant if $\mathcal{R}_i$ and $\mathcal{R}_j$ are neighboring rows
and $\mathcal{C}_k$, and $\mathcal{C}_l$ are neighboring columns,
because no straight lines exist to directly transmit stress to
infinite distance. Thus, as shown in Fig.~\ref{fig:regvsgen}, if a
generic square lattice is rigid, the corresponding regular square
lattice with the same configuration of braces must also be rigid,
but the converse is not true.
In the following, we show that due to the difference discussed above, the generic square lattice does not have any states of self stress until the bulk of the lattice is already rigid.

\begin{figure}[h]
\centering
\includegraphics[width=.5\textwidth]{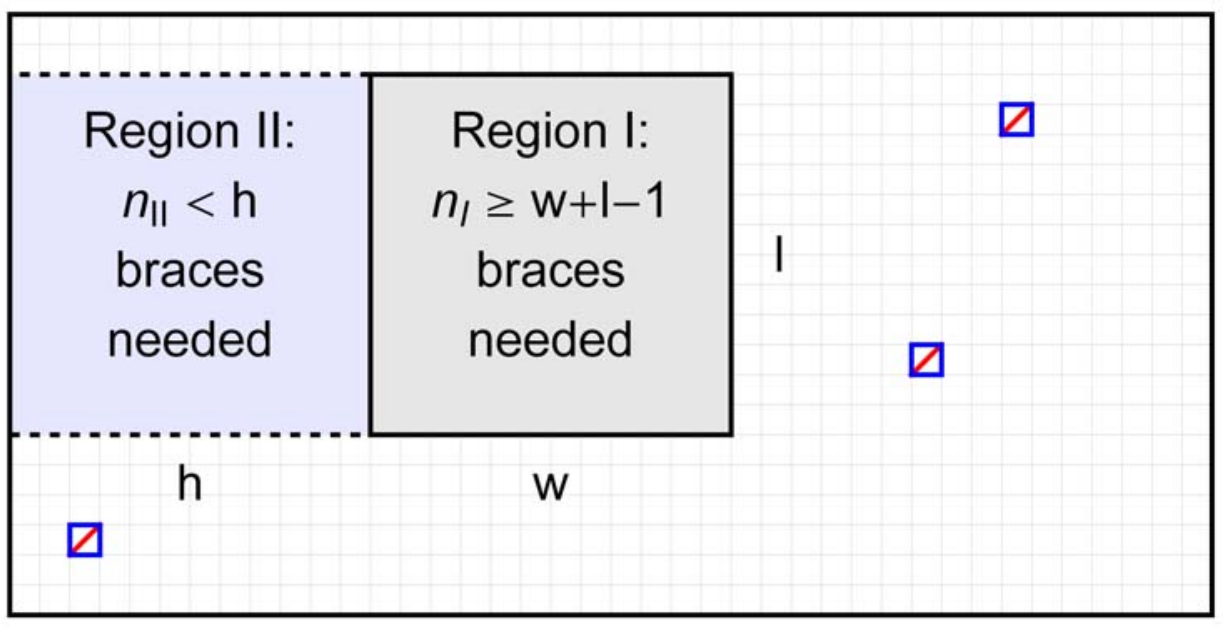}
\caption{
Consider the possibility of Region I, measuring $\ll$  by  $\lw$ plaquettes, becoming an isolated rigid region in a large generic square lattice. This would require $\ll + \lw - 1$ independent braces in Region I. However, Region II would then experience $\ll - 1$ constraints from its shared edge with Region I, and so would require only $h$ additional independent braces placed in its interior to be made rigid. For large systems, this occurs with finite probability only for $h \sim \mathcal{O}(1)$. 
Thus, as discussed in the text, rigid regions first form with nearly $\nummax = 2 \side - 3$ braces, and such regions span the entire system except possibly for a few rows and columns near the edge.
}
\label{fig:ririi}
\end{figure}

Consider Region I, a rectangular region with length $\ll$ greater than or equal to its width $\lw$, as depicted in Fig.~\ref{fig:ririi}. The region, including bonds and vertices on its boundaries
but not bonds connecting this region to neighboring vertices
, has $2 (\ll + 1) (\lw + 1)$ degrees of freedom and $ 2 \ll \lw + \ll
+ \lw$ constraints, so that $\lw + \ll -1 $ \emph{independent} braces are needed to eliminate the floppy modes of Region I. We now ask what the probability is that Region I is an \emph{isolated} rigid region.
Since braces in the generic lattice can't render distant plaquettes
rigid except by also rendering intervening ones rigid as well, only
braces within Region I itself can contribute to it becoming an isolated rigid region.
 Thus, a necessary condition for Region I to be rigid and isolated is that it contain at least $\lw + \ll -1 $ braces, which it does with probability
\begin{eqnarray}
\label{eq:lwprob}
\sum_{j = \lw + \ll - 1}^{\lw \ll} \left( \begin{array}{c} \lw \ll \\ j \end{array} \right) \pbrace^j \left(1-\pbrace\right)^{\lw \ll -j},
\end{eqnarray}
where $\pbrace = \numbrace / \left(\side-1\right)^2 $ is the probability of a brace being placed on a plaquette and 
$  \left( \begin{array}{c} \lw \ll \\ j \end{array} \right)$
 is a binomial coefficient.

\begin{figure}[h]
\centering
\includegraphics[width=.45\textwidth]{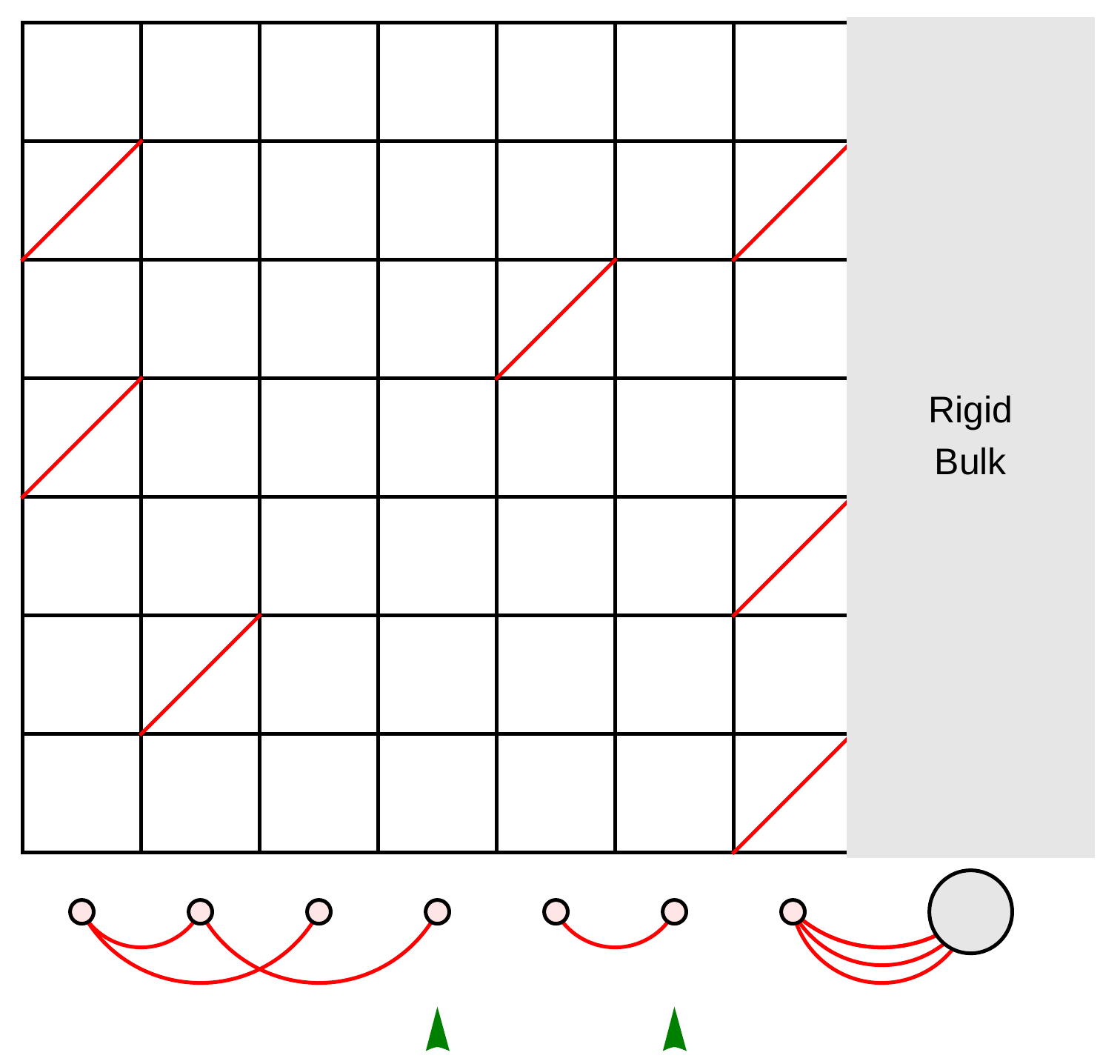}
\caption{
A floppy region on the left edge of a generic square lattice with a rigid bulk. For visual clarity, we show only a few rows and do not depict the generic displacements of vertices.
There are seven random braces in seven columns, but because of their
distribution, the edge is not rigid.
Counting from the outer edge and treating each column as a single
vertex in a graph, a brace in a column links it to the next column not
already part of the rigid cluster or to the bulk, as depicted in the
graph below the main diagram. Because the fourth and sixth columns
would require additional braces to connect the edge to the bulk, these
two columns are said to contain \emph{edge modes}, indicated by green
arrowheads. These two modes make the first six columns floppy, while
the seventh has become part of the rigid bulk. As discussed in the
text, these edge modes play an important role in the onset of
rigidity.}
\label{fig:emc}
\end{figure}

For $\lw = \ll = 2$, this probability is $\mathcal{O}(\pbrace^3)$, with larger regions higher order in $\pbrace$.
As we will see, the bulk of the system becomes rigid near the Maxwell point, when $\pbrace \sim 1/\side$, so the probability of a small isolated region occurring \emph{anywhere} in the system is only $\mathcal{O}(L^{-1})$ and vanishes for large systems. The one exception, $2\times1$ regions, occur with finite frequency but at least three braces are required to render another plaquette rigid.
This nonexistence of small rigid regions is confirmed by simulation results and permits the use of Maxwell counting in much of our analysis.  

In contrast to small regions, larger ones require a lower density of
braces $(\lw+\ll-1)/\lw\ll$ and have more ways to distribute those braces,
generating a large combinatorial factor in Eq.~(\ref{eq:lwprob}). This
suggests that large rigid regions become possible before small ones
and indeed, when $\ll$ is large, the central limit theorem applies, and the probability of having exactly $j$ braces in Region I becomes
\begin{eqnarray}
\label{eq:cltI}
\frac{1}{\sqrt{2 \pi \lw \ll \pbrace (1 - \pbrace)}} \exp \left( - \frac{\left(j - \lw \ll \pbrace\right)^2}{2 \lw \ll \pbrace (1-\pbrace)}\right),
\end{eqnarray}
so that as $\pbrace$ approaches $(\lw + \ll - 1) / (\lw \ll)$ Region I may become rigid. This occurs first for the largest regions, so it is clear already that the first rigid region to appear will cover much of the system. However, for Region I to be not only rigid but isolated Region II must remain floppy. 

When Region I is rigid, a single brace in the right column of Region II renders that entire column rigid. Such a braced plaquette, along with the rigid plaquettes
of Region I immediately to its right,
would mean that the plaquette above (or below) it would be fully constrained and rigid. 
Thus, because of the rigid edge this column, which would otherwise have $\ll$ degrees of freedom, has only one. Because of this, once Region I is rigid Region II needs only an additional $\lh$ independent braces, rather than $\ll + \lh - 1$, to be rigid as well.

On average, Region II contains at least $2 \lh$ braces, twice as many as would make it rigid, with a standard deviation in its brace number proportionate to $\sqrt{\lh}$. Thus, when Region I is rigid Region II contains sufficient braces to make it rigid as well \emph{unless} possibly its width $\lh$ is $\mathcal{O}(1)$.
Thus, when rigid regions appear at or near the Maxwell point they fill
the entire system with the possible exception of a few rows and
columns on the edges. Indeed, as depicted in
Fig.~\ref{fig:rigid_component}, as braces are added in simulation no rigid regions form until a \emph{single brace} renders the bulk of the system rigid, leaving only a small, random number of floppy rows/columns on the edges of the system in a first-order-like transition transition. We say then that the system has a rigid bulk, and we now characterize its edge modes.

\subsection{Edge modes}

We now develop a theory to describe the number of floppy modes, which
we call \emph{edge modes}, that are present on an edge when the bulk
is rigid. We say that the $m$ columns on the edge of a system have
\emph{minimal edge rigidity} if the braces present in them would
render them rigid but unstressed if the rest of the system were rigid.
Consider the first column along the left edge of the system. If the
columns to its right were rigid then a single brace would make the
entire column rigid since a rigid plaquette combined with two from the
rigid region to the right will also make the plaquettes above and
below it rigid, as in Fig.~\ref{fig:regvsgen}b. The column has
$\side-1$ plaquettes that can be sheared, but the $\side-2$ vertices (not
counting those on the edge of the system) it shares with the rigid
region couple the plaquette modes together, so that there is only one
independent floppy mode associated with this column. A single random
brace gives the first column minimal edge rigidity. Similarly, the
first two columns could be given minimal edge rigidity if two braces
were placed in the first column or if one were placed in both of the
first two columns. On the other hand, if two braces were placed in the
second column and zero in the first then the second column could be
stressed (if the third column were already rigid) while the first
would necessarily remain floppy. In general, minimal edge rigidity requires that
the $m$ columns contain exactly $m$ braces and that there isn't any
set of them connected to the bulk that contains more braces than
columns. That is, $m$ columns on the edge have minimal edge rigidity
if they contain $m$ braces distributed so that
\begin{eqnarray}
\label{eq:minrig}
\sum_{k=m-j+1}^{m} n_k \le j \textrm{ for all } j =  1,2, \ldots, m,
\end{eqnarray}
where $n_i$ is the number of braces in the $i^\textrm{th}$ column.

Consider a column such as column four in Fig.~\ref{fig:emc}. Adding a single brace to that column would give the first four columns minimal edge rigidity, so we say that that column \emph{contains an edge mode}. If we also added a brace to the sixth column, that would give the first six columns minimal edge rigidity, so we say that it too contains an edge mode. 
Generally, the requirement for an edge mode to be contained in the $\mth$ column counting inward either from the edge or from another column with a floppy edge mode is for the relations of Eq.~(\ref{eq:minrig}) to be satisfied with strict inequality. That is, the column containing the edge mode must have zero random braces, it and its left neighbor combined have one or fewer, etc.
When the bulk is rigid, the number of floppy modes associated with an edge is equal to the number of columns containing floppy edge modes as defined this way. When the bulk is not rigid, the true number of floppy modes is generally greater, since rigid regions encourage rigidity around them, as discussed above.

We now wish to determine the statistics of these edge modes. The probability, $\probm$, that the $\mth$ column contains the first edge mode is simply the probability that the first $m-1$ columns are minimally rigid and the $\mth$ column itself has no random braces. For a fixed number of random braces, this is simply the fraction of distributions of $\numbrace$ braces to the left and right of the $\mth$ column consistent with this condition, which can be expressed in terms of combinatorial factors as
\begin{widetext}
\begin{eqnarray}
\label{eq:pmeq2}
\probm=
\frac{\left( \begin{array}{c} (\side-1)(\side-m-1) \\ \numbrace-m+1 \end{array} \right) \left( \begin{array}{c} (\side-1)m \\ m-1 \end{array} \right)
}
{
\left( \begin{array}{c} (\side-1)^2 \\ \numbrace \end{array} \right)
}
\sum_{\{n_i\}_{i=1}^m}' \prod_{i=1}^{m} \left( \begin{array}{c} (\side-1) \\ n_i \end{array} \right)
,
\end{eqnarray}
\end{widetext}
where the sum is over only those brace distributions $\{n_i\}_{i=1}^m$ consistent with an edge mode being present in the $\mth$ column. For large systems the number of braces per column follows a Poisson distribution with a mean value  $\bpc \equiv \numbrace/(\side-1)$ braces per column, so that
\begin{eqnarray}
\label{eq:pmeq3}
\probm = e^{-m c} c^{m-1} \sum_{\{n_i\}_{i=1}^m}' \prod_{i=1}^{m} \frac{1}{n_i!}.
\end{eqnarray}
One can add a brace to any of $m$ columns in order to make the edge
mode minimally rigid, so that the combinatorial factor is $m^{-1}$
times the equivalent for a minimally rigid set of braces. The
minimally rigid set of braces on $m$ columns can be mapped onto the
set of spanning trees of a graph of $m$ distinguishable edges by
noting that, counting from the edge, each brace connects the column it
is in to the first column that is not already in the rigid cluster
(see Fig.~\ref{fig:emc}). Thus, applying Cayley's formula for the
number of spanning trees of a complete graph~\cite{Cayley1889},
\begin{eqnarray}
\label{eq:pmeqfinal}
\probm = e^{-m c} c^{m-1} \frac{m^{(m-2)}}{\left(m-1\right)!}.
\end{eqnarray}
$\probm$ is only physically meaningful when the rigid bulk is present, near or above $\numbrace = \nummax$. Then, $\probm$ quickly falls as $m$ increases, and even in large systems only a few columns at each edge are potentially floppy. Then, the probability $\probrigidedge$ that the edge will be rigid when the bulk is rigid is
\begin{eqnarray}
\label{eq:probrigidedge}
\probrigidedge = 1 - \sum_{m=1}^\side \probm(\numbrace),
\end{eqnarray}
where the sum quickly converges for $\numbrace \gtrsim \nummax/2$, so that columns far from the edge don't contain edge modes.

Thus far, we have considered only the \emph{first} edge mode on an
edge. However, an edge may contain two or more modes, as in
Fig.~\ref{fig:emc}. Once the first edge mode has been identified, the
conditions under which a second appears $m$ columns inward are simply
those of Eq.~(\ref{eq:minrig}), applied to the $m$ columns to the right of the first edge mode, rather than to the first $m$ columns counting from the outer edge. Thus, an edge contains $\numem$ edge modes with probability $\probrigidedge (1-\probrigidedge)^{\numem}$ and
\begin{eqnarray}
\label{eq:numem}
\langle \numem \rangle = \frac{1-\probrigidedge}{\probrigidedge}.
\end{eqnarray}
Although we have relied on the concept of a rigid bulk to describe these edge modes, it is the edge modes themselves that determine when the bulk becomes rigid. Consider a system with a total number of edge modes $\sum \numem$ which leave $\numcol$ columns and rows floppy. Since the floppy edges do not support states of self stress, Maxwell counting indicates that they contain $\numcol - \sum \numem$ random braces and that the remaining bulk comprises an area originally containing $\nummax - \numcol$ floppy degrees of freedom. This leads to the criterion for bulk rigidity
\begin{eqnarray}
\label{eq:bulkcriterion}
\nummax - \numbrace \le \sum_{\textrm{edges}} \numem.
\end{eqnarray}
That is, the bulk is rigid even below the Maxwell point so long as the needed floppy modes can all be found at the edge. The bulk is always rigid above the Maxwell point, since no states of self stress occur without a rigid bulk.
When the bulk first becomes rigid the above relationship is satisfied with equality. Since even in large systems only a few edge modes occur with substantial probability the bulk becomes rigid either at the Maxwell point or only a few braces below. This is a first order transition in which a single brace makes all but perhaps $\mathcal{O}(\side)$ plaquettes rigid.
Then, as additional braces are added to the system, each one either eliminates an edge mode or generates a state of self stress within the bulk. 

\subsection{Rigidity statistics}

The probability $\probbulk$ that the bulk of the system is rigid is simply the probability that the condition of Eq.~(\ref{eq:bulkcriterion}) is met:
\begin{eqnarray}
\label{eq:probbulk}
\probbulk(\numbrace) = 1 - \textrm{Pr}\left(\sum_{\textrm{edges}} \numem < \nummax - \numbrace\ \right).
\end{eqnarray}
For large systems, the corners where row and column modes
meet are negligible and the statistics of the modes on different edges
follow independently from Eq.~(\ref{eq:pmeqfinal}). This probability
$\probbulk$ is plotted in Fig.~\ref{fig:bulk_rigidity_probability}, in quantitative agreement with simulation.

This bulk rigidity probability also determines the mean number of floppy modes present. When the bulk is not rigid, the number of floppy modes follows from Maxwell counting. When, on the other hand, the bulk is rigid the edge modes of Eq.~(\ref{eq:numem}) are the only modes present so that generally
\begin{eqnarray}
\label{eq:expf}
\avgfloppy = \left(1-\probbulk\right)\left(\nummax - \numbrace\right)+ \probbulk \frac{4 \probrigidedge}{1-\probrigidedge}.
\end{eqnarray}
Well below the Maxwell point each brace eliminates a floppy mode, as indicated by the first term. At or above $\nummax$, only the edge modes from the second term are present. Slightly below the Maxwell point the system may or may not have a rigid bulk, and both edge modes and bulk floppy modes contribute to different lattice realizations seen in simulation.

Separately from the rigidity of the bulk, there is the probability $\probrigid$ that the system is entirely rigid, without even edge modes. This can occur only for $\numbrace \ge \nummax$, and requires simply that all four edges be rigid as described in Eq.~(\ref{eq:probrigidedge}). Thus,
\begin{eqnarray}
\label{eq:probrigid}
\probrigid =     \left\{ \begin{array}{lr}
     0 &\textrm{  for  } \numbrace < \nummax  \\
     R^4 &\textrm{  for  } \numbrace \ge \nummax
     \end{array}
   \right.
\end{eqnarray}
At the Maxwell point, this has finite probability $\approx 0.403$, the
probability that no edge modes are present. Unlike bulk rigidity,
which is achieved within a few braces of the Maxwell point, total
rigidity generally requires $\mathcal{O}(\side)$ additional braces, since each brace is much more likely to fall in the bulk than to eliminate an edge mode.

The picture we have developed is for the generic square lattice, but applies without substantial modification to the generic kagome lattice with a hexagonal geometry. For the kagome, there are six edges and three principle directions, but the edge modes on each edge are eliminated by random braces in much the same way as the square lattice. Unlike the square lattice, where every column has $\side-1$ sites to place random braces, the $\mth$ ``column'' from an edge in the kagome lattice has $\side + m - 2$ sites, but since only the first few columns can contain edge modes this does not affect the behavior of large systems.

This analytic theory of edge modes thus predicts the probability of the rigid bulk, the probability of total rigidity, and the average number of floppy modes, as shown respectively in Fig.~\ref{fig:bulk_rigidity_probability}, Fig.~\ref{fig:pr} and Fig.~\ref{fig:floppy modes}. Using no free parameters, it achieves quantitative agreement with the behavior of the simulations of generic square and kagome lattices below, above, and precisely at the rigidity transition. 

Were we instead to work in an ensemble with fixed brace \emph{probability} $p$, $\numbrace$ would become a random variable with, to leading order for large lattices, mean $p \side^2$ and standard deviation $\side \sqrt{p}$. This would lead to a rigidity transition at $\rigidthreshold = 2 / \side$ in which number fluctuations would smooth out the transition that otherwise occurs via a single brace to one that takes place over a range of probabilities $\Delta p \sim L^{-3/2}$.

\section{Conclusion and Discussion}\label{SEC:CD}

We have elucidated the rich phenomenology of rigidity transitions
in regular and generic isostatic lattices with $\numbrace$ added braces.  We now summarize our main findings.

Regular
lattices become rigid after approximately $\mathcal{O}(\side\ln \side)$ braces are
added, in accordance with ``coupon-collector'' heuristics.
However, they first undergo a transition at $\mathcal{O}(\side)$.  In the
regular square lattice, this transition is weak and has a discontinuity
in the third derivative of the average number of floppy modes
$\avgfloppy$.  
The regular kagome lattice appears to form a giant
rigid cluster at $\mathcal{O}(\side)$ via a first-order transition---
$\avgfloppy/\side$ then decays exponentially in
$\numbrace-\nummax$ until the rigidity transition occurs.

In generic lattices, the nature of the rigidity transition is quite
different. No extended rigid regions exist in such a lattice until a
single brace renders the entire bulk of the system rigid at or a few braces before the Maxwell point. Once the bulk is rigid, floppy edge modes may exist and persist well above the Maxwell point even as self stresses proliferate in the bulk.

In both types of systems, despite the fact that the rigidity
transitions are first order, the transition probabilities scale as
$1/\side$ (to lowest order). This determines at fixed $p$ a critical
system size which diverges like $1/p$.

Below we point out some connections between the bracing percolation
problem to other work and suggest some directions for future work.

A previous study of braced generic square lattices~\cite{Moukarzel1999} attached the system to rigid bars along diagonals, preventing the appearance of edge modes and thereby altering the nature of the transition. How, then, do the boundary conditions and the shape of the boundary influence the edge modes and the rigidity transition? Fixing certain boundaries may lead transitions in which edge self stresses rather than floppy modes control the behavior.

The transition we observe appears to be very closely related to the
rigidity transition on the Erd\H{o}s-R\'enyi model on the
complete
graph~\cite{moukarzelfield,Kasiviswanathan,rivoire,Barre2014}. Just as
in our braced isostatic lattices, the systems exhibit the sudden
appearance of a giant rigid cluster when the number of edges is
$\mathcal{O}(1/\numsites)$. Our
arguments for the nonexistence of small rigid clusters in
Sec.~\ref{SEC:TheoryGeneric} are similar to
those made in Refs.~\cite{Theran09,Kasiviswanathan} and likely can be
made rigorous along similar lines. Several authors
\cite{moukarzelfield,Barre2014} have considered the
problem of rigidity percolation on complete graphs with an additional
``applied field'' of random additional pin and slider constraints to a fixed
background and have found true
critical and tricritical behavior in the formation of a giant rigid
cluster.  It would be interesting to see whether addition of such
constraints also induce similar phenomena in the bracing percolation problem.

In generic isostatic lattices, we find that the lattice has no self stresses until a compact rigid bulk occurs very close to the Maxwell point.  This is strikingly similar to the observation in jamming that the whole system becomes jammed at the Maxwell point of coordination without self stress~\cite{Liu2010}.  Another similarity between generic isostatic lattices and jamming is that the addition of a single brace above this point renders the system globally stressed~\cite{Ellenbroek2014}.  In addition, the two systems show the same scaling of diverging length near isostaticity which agrees with the cutting argument from Ref.~\cite{Wyart2005a}. These similarities may indicate a deeper relation.  In this sense, the generic isostatic lattices are closer to jamming than either regular isostatic lattices or diluted generic triangular lattices, because the latter two can develop self stresses before rigidity percolation.

In this paper we took the point of view of changing the density of
braces while fixing the system size. One can also frame the
results by instead imagining what happens if the density of braces is fixed and the system size
is changed.  In particular, one could imagine cutting out from a large
system a sample with linear size $\side$ and considering the rigidity
properties of this sample.  
From our results, we see that for generic systems, provided
that $\side$ is sufficiently small so that $\numbrace\leq\nummax$, the system consists of many small rigid regions,
and when $\side$ is large enough that $\numbrace\geq\nummax$ the
bulk of the system rigidifies.
In Ref.~\cite{goodrich}, the authors consider networks arising from jammed packings and use the system size
$\side$ at which the bulk of the system rigidifies to {\em
define} a rigidity length scale $l^*$.  
It would be very interesting to pursue further connections with
jamming---e.g. whether the behavior of the rigidity length scale can
be understood as arising from a first order rigidity transition as in
the randomly braced lattices.

Bootstrap percolation/$k$-core percolation
\cite{aizenmanlebowitz,holroyd,Schwarz2006} and kinetically-constrained models
\cite{kineticallyconstrainedreview} are other combinatorial models on
graphs which have been used to study jamming transitions.  They have
some similar features; in the closest-related lattice bootstrap percolation
models, where lattice sites are deemed active with fixed probability
\cite{aizenmanlebowitz,holroyd}, as the system size grows the
critical occupation probability goes to zero.  In those models, this has been interpreted as a kind of metastability -- the idea being that
there is a size above which the system is likely to contain a
``critical droplet'' which causes the entire system to be jammed
\cite{aizenmanlebowitz}. In a $k$-core problem on the Bethe lattice,
a mixed first-and-second-order transition was observed, with the
fraction of sites in the spanning cluster undergoing a discontinuous
jump followed by critical scaling \cite{Schwarz2006}.

The mechanism of rigidity percolation in generic braced isostatic lattices
has some features of both of these transitions.  In the systems we study, the critical length
scale arises from the difference in the scaling with system size
between the number of floppy modes
coming from the free boundary and the number of added braces when the
density is held fixed.  Nonetheless there may still be some
metastability phenomena. If braces are added at random until the system becomes rigid and then removed at random one by one, then due to
the random distribution of self stresses, the system is likely to lose
rigidity with a different number of braces than that with which it gained rigidity.
However, our results show that the width of this metastability window
should be quite small, approximately $\mathcal{O}(1)$ in the generic systems. We
do not observe critical exponents above the first-order like jump
above $\rigidthreshold$ in our systems, as in the $k$-core problem of
Ref.~\cite{Schwarz2006}, however we can identify a diverging length
scale from the system size dependence of $\rigidthreshold\sim1/\side$. 
The connections between bootstrap / $k$-core percolation models and
bracing percolation deserve to be further studied.
One can also ask whether braced isostatic systems exhibit ``jamming by
shape'' as some kinetically-constrained models do \cite{teomyshokef}.

Because one can continuously tune a lattice between regular and generic by small perturbations of lattice sites positions, it will be interesting to examine how some floppy modes in the regular isostatic lattices are lifted to finite energy, whereas some keep being floppy modes, as well as how modes crossover from extended to localized.

For the generic bracing percolation systems we consider, the shape distribution
of the eventual giant rigid cluster can be computed fairly easily
because rigid clusters must be either rectangular or hexagonal. It
would be interesting to compare this to the average shape of the
typical large rigid cluster in the jammed packings of Ref.\
\cite{goodrich}.
While the rigid clusters seem to have a simple shape, the plots in
Fig.~\ref{fig:rigid_component} suggest questions about the
distribution and size of {\em stressed} regions (yellow bonds).  The
stressed regions have significance for the robustness of the systems,
as they consist of the bonds that can be removed without making the
system floppy.

\section{Acknowledgments}

BGC thanks Louis Theran for illuminating discussions.
BGC was supported by NSF DMR05-47230 as well as by the Foundation for Fundamental Research on Matter
(FOM), which is part of the Netherlands Organisation for Scientific
Research (NWO).

\appendix 

\section{Braced rigidity matrix for the braced regular kagome
lattice}

In this appendix, we derive a simplified rigidity matrix for next
nearest neighbor bonds on the regular kagome lattice, which we call
the braced rigidity matrix.  This matrix
representation is used in the rank calculations in the numerical
results of Section \ref{SEC:Simulation}.

The usual rigidity matrix keeps track of all $2\numsites$ possible
displacements of the $\numsites$ points in a spring network and each row of
the matrix expresses how these displacements are coupled to each other
by each spring in the system.  The rigidity matrix thus is $\nc \times 2 \numsites$.

\begin{figure}
\centering
\includegraphics[width=0.5\textwidth]{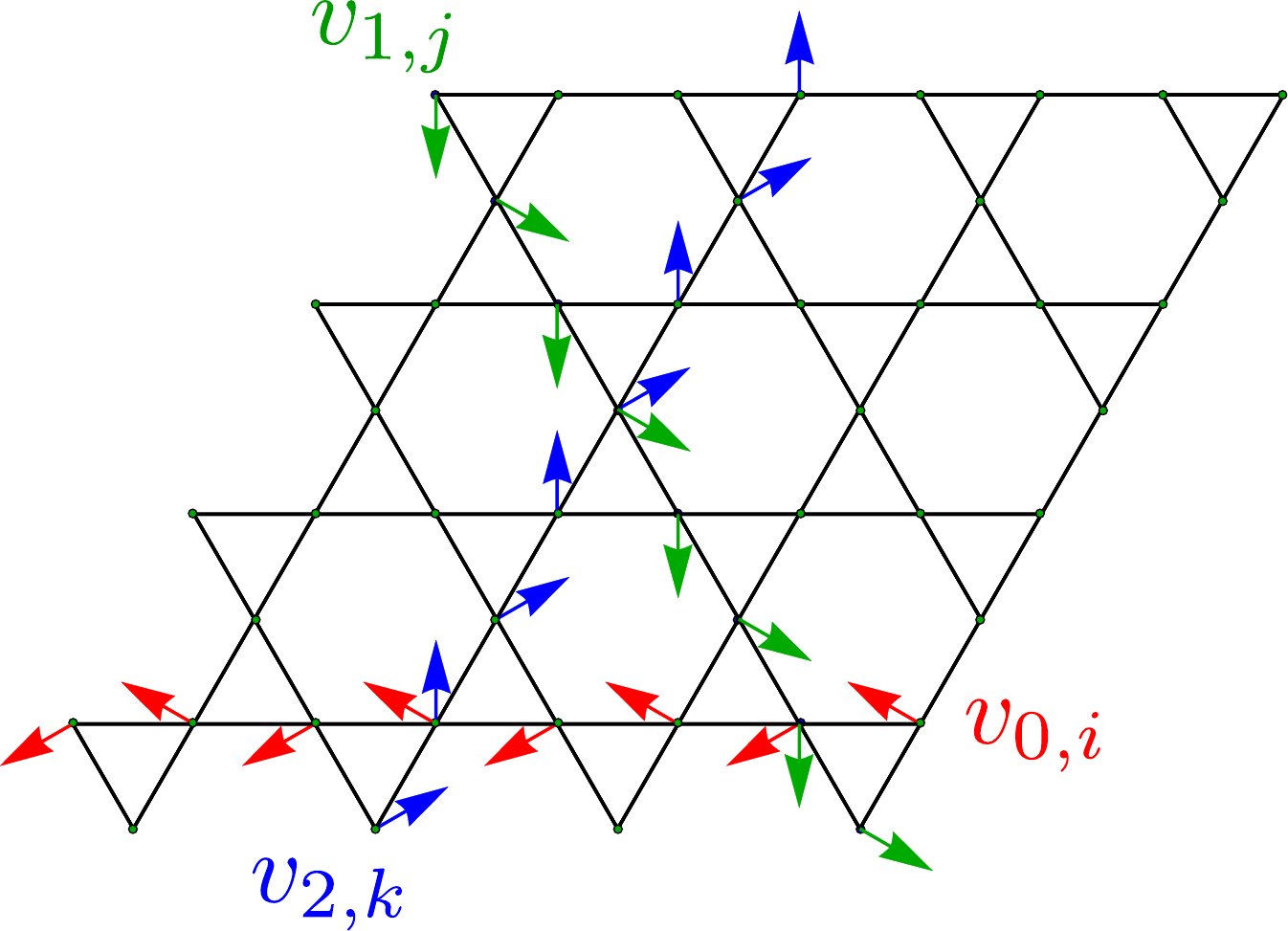}
\caption{Three line modes in a portion of the regular kagome lattice.
The line modes supported on horizontal lines (red) are denoted $v_{0,i}$,
those on lines with angle $2\pi/3$ (green) are denoted $v_{1,j}$ and those on
lines with angle $4\pi/3$ (blue) are denoted $v_{2,k}$. On a hexagon,
the indices $i,j,k$ run from 1 to $2\side$.}
\label{fig:kagomelinemodes}
\end{figure}

The braced rigidity matrix instead uses
only the degrees of freedom that a regular kagome lattice allows (arising
from modes localized on 3 families of lines, see Fig.\
\ref{fig:kagomelinemodes}). We now consider how each brace couples these degrees of freedom together.

The three families of lines are at
angles of $0$, $2\pi/3$ and $4\pi/3$ relative to the $x$-axis.  
The floppy mode localized on a horizontal straight line
$l$ in the kagome lattice has an infinitesimal displacement on each
vertex equal to $\left(\frac{\sqrt{3}}{2},\pm\frac{1}{2}\right)$ (with the signs
alternating and chosen so that the displacement at a vertex is always perpendicular to the 
non-horizontal line intersecting $l$ there). Other line-localized modes can
be generated similarly.

We now introduce some notation. We
denote the $2\numsites$-dimensional vectors corresponding to the modes
supported on lines in these directions as $v_{0,i}$, $v_{0,j}$, and
$v_{0,k}$, respectively, where the indices $i,j,k$ label the
specific line with the angle specified by the first index.  Note that
for a hexagonal system, there are $2\side$ lines running in each
direction.  It is straightforward to check that
the $6\side$ modes $v_{0,i}$, $v_{1,j}$, $v_{2,k}$ span
the space of zero modes (floppy modes as well as translations and rotations)
of the regular kagome hexagon with no braces.

\begin{figure}
\centering
\includegraphics[width=0.5\textwidth]{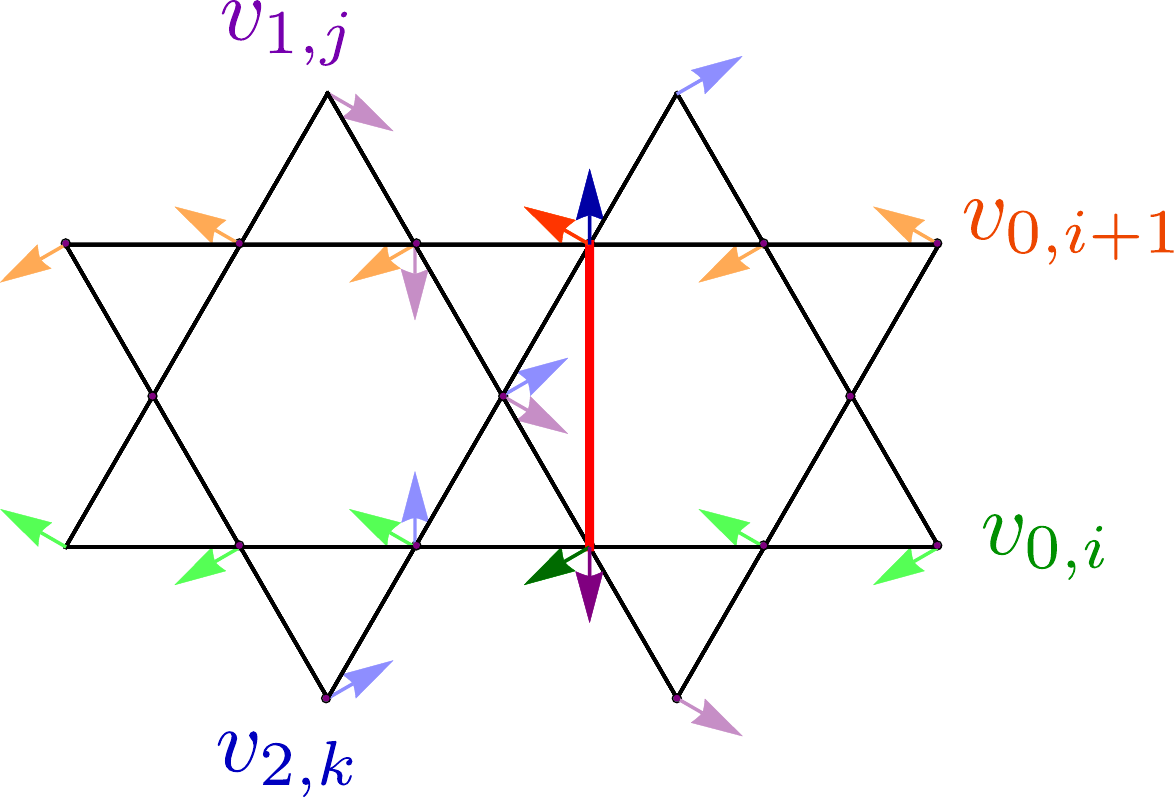}
\caption{The constraint from a brace (red) on the coefficients of four line modes
$v_{0,i}$ (green), $v_{0,i+1}$ (orange), $v_{1,j}$ (purple) and
$v_{2,k}$ (blue).  Suppose that the coefficients are
$c_{0,i}$, $c_{0,i+1}$, $c_{1,j}$, $c_{2,k}$. Then the brace is
unstretched if the $y$-component of the velocity at the upper particle
is equal to the $y$-component of the velocity at the lower particle:
thus $-c_{0,i}-2c_{1,j}=c_{0,i+1}+2c_{2,k}$, or equivalently, 
$c_{0,i}+c_{0,i+1}+2c_{1,j}+2c_{2,k}=0$.}
\label{fig:1221}
\end{figure}

Consider now a brace, which couples the motions of two particles
together.  The constraint imposed by requiring this brace not to
stretch to linear order is that the difference in displacements of the two 
particles must be perpendicular to the direction of the bond.
Since each particle is at the intersection of two lines,
there are only two modes which contribute to the motion of any
particle.  Thus there are four line modes which are constrained by the
brace, which are always in a configuration like that in Fig.\
\ref{fig:1221} or some rotation thereof.  Let us suppose these four
modes are $v_{0,i}$, $v_{0,i+1}$, $v_{1,j}$ and $v_{2,k}$.
If the coefficients of these four modes in some motion are
$c_{0,i}$, $c_{0,i+1}$, $c_{1,j}$, $c_{2,k}$. Then the brace is
unstretched if the $y$-component of the velocity at the upper particle
is equal to the $y$-component of the velocity at the lower particle:
\begin{align}
-c_{0,i}-2c_{1,j}&=c_{0,i+1}+2c_{2,k}
\end{align}
\noindent or equivalently
\begin{align}
c_{0,i}+c_{0,i+1}+2c_{1,j}+2c_{2,k}&=0.
\end{align}
We have identical linear equations for the other braces, though the
indices are different. Note that all of
these have integer coefficients.  We combine these equations together 
for every brace and the resulting integer matrix is $\numbrace$ by
$6\side$.  This is the braced rigidity matrix.

\section{Floppy modes of the regular square lattice}

In this appendix, we compute the number of floppy modes of the randomly
braced regular square lattice.  This is done by exploiting the map to
a random bipartite graph model (described in Sec.\
\ref{SEC:TheoryRegular}, see also Fig.\ \ref{fig:fmc}) and adapting the results of Engel et al.~\cite{engel} who use the 
Fortuin-Kasteleyn cluster expansion to write the expected number of connected
components with a certain $q$-weighted distribution as the $q\rightarrow1$ limit of the magnetization in a
Potts model. 

The complete bipartite graph $K_{\side,\side}$ is the graph with two
partitions of $\side$ vertices
$P_1,P_2$, such that every vertex in $P_1$ is adjacent 
to every vertex in $P_2$ (and vice versa), but is not adjacent to any
vertex in $P_1$, and similarly, vertices in $P_2$ are not adjacent to
any vertices in $P_2$.  
Note that for notational simplicity in this section we
work with a $(\side+1)\times(\side+1)$ square grid, so that 
the vertices in $P_1$ correspond to the $\side$ adjacent pairs of rows, and the vertices in $P_2$
correspond to the $\side$ adjacent pairs of columns.

Let $G(K_{\side,\side},\gamma)$ be a random graph on the set of vertices of 
$K_{\side,\side}$ where each of the $\side^2$ edges of $K_{\side,\side}$ is present with
probability $p=\gamma/\side$, independently. This gives a bipartite Erd\H{o}s-R\'enyi
type model. Engel et al.\ related the Erd\H{o}s-R\'enyi model on the
(ordinary) complete graph to the Potts model by studying a 
probability distribution on random graphs which is {\em biased}
towards having either more or fewer connected components depending on
a new parameter $q$. Taking the limit $q\rightarrow1$ yields
results relevant for the unbiased distribution.
Below we adapt their work to $K_{\side,\side}$. 

We first define the Potts model on this graph. Each vertex of
$K_{\side,\side}$
carries a spin variable $\sigma_i$ which can take any of $q$ values,
that is $\sigma=0,1,\dots,q-1$. For later convenience, the
spin variables on vertices in $P_1$ will be called $\sigma_i$ for
$i=1$ to $\side$ and the spin variables on vertices in $P_2$ will be
called $\tau_j$ for $j=1$ to $\side$.  The energy function of a spin
configuration (at zero field, which suffices for our purposes in this
section) is then
\begin{align}
E(\{\sigma_i,\tau_j\})=&-\frac{1}{2\side}\sum_{i=1}^\side\sum_{j=1}^\side\sigma_i\tau_j.
\end{align}

The partition function is
\begin{align}
\mathcal{Z}(\beta,q,\side)=&\sum_{\{\sigma_i,\tau_j\}}\exp(-\beta
E(\{\sigma_i,\tau_j\})),\label{eq:pottspartition}
\end{align}
where we sum over all $q^{2\side}$ possible spin configurations. The
free energy (per site) in the thermodynamic limit
$(\side\rightarrow\infty)$ is
\begin{align}
f(\beta,q)=&-\lim_{\side\rightarrow\infty}\frac{1}{2\beta
\side}\ln\mathcal{Z}(\beta,q,\side).
\end{align}
If $c(\gamma)$ is the typical number of components per
vertex in a graph in the Erd\H{o}s-R\'enyi model on $K_{\side,\side}$ with
parameter $p=\gamma/\side$ then 
to leading order in $\side$, the results of \cite{engel} show that
\begin{align}\label{eq:numcomponents}
c(\gamma)&=(2\gamma)\left.\frac{\partial f(2\gamma,q)}{\partial
q}\right|_{q=1}.
\end{align}

We may calculate the free energy of the Potts
model on $K_{\side,\side}$ using a mean-field ansatz, see e.g.\ \cite{wupotts}. 
We begin by introducing the following $2q$ order parameters (``magnetizations'' of
each type of spin in $P_1$ and $P_2$):
\begin{align}
m_k=&\frac{1}{\side}\sum_{i=1}^\side\delta(\sigma_i,k)\\
n_k=&\frac{1}{\side}\sum_{j=1}^\side\delta(\tau_j,k).
\end{align}
Thus $m_k$ is the fraction of spins in $P_1$ which are in spin state
$k$ for $k=0,1,\dots,q-1$, and similarly $n_k$ is the fraction of
spins in $P_2$ in state $k$, and hence $\sum_km_k=\sum_kn_k=1$. 
In terms of these variables (and neglecting fluctuations), our energy
function $E(\cdot)$ becomes
\begin{align}
E(\{m_k,n_k\})=&-\frac{N}{2}\sum_{k=0}^{q-1}m_kn_k. 
\end{align}
It turns out that we get a simplification here because the sizes
of $P_1$ and $P_2$ are the same.  In particular, we shall see that the
free energy is very nearly the same as that of the usual mean-field
Potts model on a complete graph.

In going from the microscopic variables $\{\sigma_i,\tau_j\}$ to the
macroscopic variables $\{m_k,n_k\}$ we get an entropy of mixing term as well:
\begin{align}
S(\{m_k,n_k\})=&-k_BN\sum_{k=0}^{q-1}\left(m_k\ln m_k+n_k\ln
n_k\right).
\end{align}

\noindent To compute the free energy (per site), we must extremize $E-TS$:
\begin{align}
\beta f(\beta,q)=&\operatorname*{extr}_{\{m_k,n_k\}}\sum_{k=0}^{q-1} \left(m_k\ln m_k+n_k\ln
n_k-\frac{\beta}{2}m_kn_k\right).
\end{align}

We apply the following ansatz, which assumes that symmetry will be broken
in the $k=0$ spin direction: 
\begin{align}
m_0=&\frac{1}{q}[1+(q-1)s]\\
m_k=&\frac{1}{q}(1-s),\qquad k=1,2,\dots,q-1\\
n_0=&\frac{1}{q}[1+(q-1)s]\\
n_k=&\frac{1}{q}(1-s),\qquad k=1,2,\dots,q-1.
\end{align}
There is now a single order parameter $0\leq s\leq 1$, the spontaneous ``magnetization'' of the Potts model.  We have:
\begin{align}
\beta f(\beta,q)=&\operatorname*{extr}_s\left\{\frac{2}{q}[1+(q-1)s]\ln\left(\frac{1}{q}[1+(q-1)s]\right)\right.\nonumber\\
&\qquad\quad-\frac{\beta}{2}\left(\frac{1}{q}[1+(q-1)s]\right)^2\nonumber\\
&\qquad\quad+\frac{2(q-1)}{q}(1-s)\ln\left(\frac{1}{q}(1-s)\right)\nonumber\\
&\qquad\quad\left.-\frac{\beta(q-1)}{2}\left(\frac{1}{q}(1-s)\right)^2\right\},
\end{align}

\noindent which simplifies to
\begin{align}
\beta f(\beta,q)=&\operatorname*{extr}_s\left\{\frac{2}{q}[1+(q-1)s]\ln\left(1+(q-1)s\right)\right.\nonumber\\
&\qquad\quad+\frac{2(q-1)}{q}(1-s)\ln\left(1-s\right)\nonumber\\
&\qquad\quad\left.-2\ln q
-\frac{\beta}{2q}\left(1+(q-1)s^2\right)\right\}.
\end{align}
This expression nearly coincides with the result for the complete
graph in Ref.\ \onlinecite{engel}. In particular, $(2\gamma)
f(2\gamma,q)$ on the complete bipartite graph is
equal to $\gamma f(\gamma,q)$ on the complete graph.
An intuitive reason for this is that the ``local neighborhood''
of every vertex in the bipartite graph looks exactly like that of a
complete graph, and the mean field assumption ensures that this is all
that matters.

Let $s_*(\beta,q)$ be the value of $s$ which extremizes the above, then
$s_*$ is the stable solution of

\begin{align}\label{eq:sstar}
e^{\beta
s_*(\beta,q)/2}=&\frac{1+(q-1)s_*(\beta,q)}{1-s_*(\beta,q)}.
\end{align}

\noindent We now specialize to $q=1$, which describes results for connectivity
percolation.  Now the order parameter $s_*$, which was the spontaneous
``magnetization'' in the Potts model, should be interpreted as the
percolation probability, i.e. the probability that a given site is
connected to the giant component \cite{wupercolation}.  Translating
further, into the language of rigidity on the regular square lattice,
$s_*$ is the probability that a given column or row shear mode is
coupled to the ``giant floppy mode''.

Our goal is now to compute the number of connected components on this
bipartite graph, which we use in Eq.\ (\ref{eq:numcomponents}) for the
number of floppy modes. The result is:
\begin{align}
c(\gamma)=&(1-s_*(2\gamma,1))\left(1-\frac{\gamma}{2}(1-s_*(2\gamma,1)\right).
\end{align}
Recall that we defined $\gamma/\side = p$; thus we have from Eq.\
(\ref{eq:sstar}) that $s_*(2p\side,1)$ satisfies
$1-s_*(2p\side,1)=e^{-p\side s_*(2p\side,1)}$.

\begin{figure}
\centering
\includegraphics[width=0.5\textwidth]{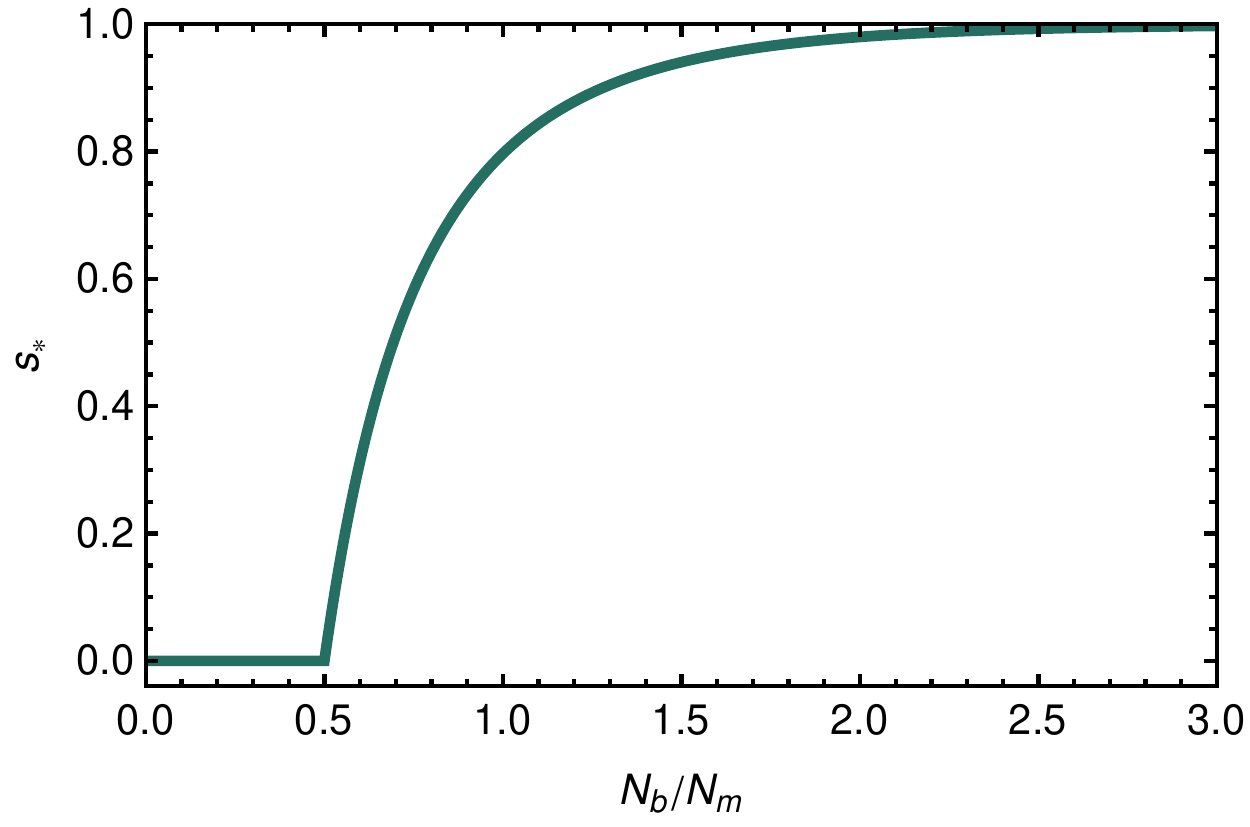}
\caption{The bipartite graph percolation probability $s_*$ as a
function of $\numbrace/\nummax$. A giant component appears
continuously at $\numbrace/\nummax=1/2$--- the singularity there is
governed by the critical exponent $\beta$, which takes the mean field
percolation value $1$.}
\label{fig:sstar}
\end{figure}

In our numerics we have been scaling the number of floppy modes by
dividing by $\side$. Here $c$ was defined as the number of connected
components {\em per vertex}, and so we divided by $2\side$ in its
definition rather than $\side$.  Hence we must multiply $c$ by two to
get $\avgfloppy/\side$.
Thus the number of floppy modes is 
\begin{align}\label{eq:solution}
\frac{\avgfloppy}{\side}=&2(1-s_*(p\side))\left(1-\frac{p\side}{2}(1-s_*(p\side))\right),
\end{align}
with $s_*(p\side)$ satisfying $1-s_*=e^{-p\side s_*}$.
The bipartite graph percolation probability $s_*$ as a function of
$\numbrace/\nummax$ (which with $\numbrace$ translating to $p\side^2$ and
$\nummax=2(\side+1)-3$ is equivalent to $p\side/2$ at large $\side$) is shown in Fig.~\ref{fig:sstar}.  Note that there is a cusp at
$\numbrace/\nummax=1/2$, and thus the appearance of the giant floppy
mode is {\em not} at the Maxwell point.  It may easily be shown from the
mean field equation for $s_*$ that there is a finite slope at the
transition, and this means that the
critical exponent $\beta$ governing the singularity there is equal to
1.

Fig.\ \ref{fig:regular_floppy_plots}(a) shows a comparison between the
prediction for the floppy modes from Eq.\ (\ref{eq:solution})
(blue line) and the number of floppy modes measured for square grids
with $N=100,200$ and 300.

\bibliography{isostaticity3}

\end{document}